\documentclass[twocolumn, switch]{article} % Method A for two-column formatting

\usepackage{preprint}
\usepackage{algorithm}
\usepackage{algorithmic}
\usepackage{amsmath, amsthm, amssymb, amsfonts}
\usepackage{enumerate}
\usepackage{enumitem}
\usepackage{subfig}
\usepackage{multirow}

\usepackage{booktabs}
\usepackage{tabularx}
\usepackage[numbers,square]{natbib}
\usepackage[utf8]{inputenc}	% allow utf-8 input
\usepackage[T1]{fontenc}	% use 8-bit T1 fonts
\usepackage{xcolor}		% colors for hyperlinks
\usepackage[colorlinks = true,
            linkcolor = purple,
            urlcolor  = blue,
            citecolor = cyan,
            anchorcolor = black]{hyperref}	% Color links to references, figures, etc.
\usepackage{booktabs} 		% professional-quality tables
\usepackage{nicefrac}		% compact symbols for 1/2, etc.
\usepackage{microtype}		% microtypography
\usepackage{lineno}		% Line numbers
\usepackage{float}			% Allows for figures within multicol

\usepackage{lipsum}		%  Filler text

\usepackage{newfloat}
\DeclareFloatingEnvironment[name={Supplementary Figure}]{suppfigure}
\usepackage{sidecap}
\sidecaptionvpos{figure}{c}

\usepackage{titlesec}
\titlespacing\section{0pt}{12pt plus 3pt minus 3pt}{1pt plus 1pt minus 1pt}
\titlespacing\subsection{0pt}{10pt plus 3pt minus 3pt}{1pt plus 1pt minus 1pt}
\titlespacing\subsubsection{0pt}{8pt plus 3pt minus 3pt}{1pt plus 1pt minus 1pt}

\title{Robust and cost-effective quantum network using Kramers-Kronig receiver}

\usepackage{eso-pic}
\usepackage{tikz}
\usepackage{xcolor}
\PassOptionsToPackage{colorlinks=true,linkcolor=gray,urlcolor=gray}{hyperref}

\newcommand{\AddMyWatermarks}{%
  \begin{tikzpicture}[remember picture, overlay]
  \end{tikzpicture}%
}

\AddToShipoutPictureBG*{\AddMyWatermarks}

\usepackage{titling}
\usepackage{orcidlink}
\usepackage{footmisc}
\newcommand{\Author}[3]{% Name, ORCID, Institution
  \textbf{#1}\textsuperscript{#2}\ %\orcidlink{#3} %
}

\author{
  \Author{Xu Liu}{1}\and,
  \Author{Tao Wang}{1,2,3,$*$}\and,
  \Author{Junpeng Zhang}{1}\and,
  \Author{Yankai Xu}{1}\and,
  \Author{Yuehan Xu}{1} \and,
  \Author{Lang Li}{1}\and,
  \Author{Peng Huang}{1,2,3}{},\and
  \textbf{and}  \Author{Guihua Zeng}{1,2,3,$\dag$}{}
}

\date{%
  \textsuperscript{1}State Key Laboratory of Photonics and Communications,	Center for Quantum Sensing and Information Processing, Shanghai Jiao Tong University, Shanghai 200240, China\\
  \textsuperscript{2}Shanghai Research Center for Quantum Sciences, Shanghai 201315, China\\
  \textsuperscript{3}Hefei National Laboratory, CAS Center for Excellence in Quantum Information and Quantum Physics, Hefei 230088, China\\
  \textsuperscript{$*$}tonystar@sjtu.edu.cn\\
  \textsuperscript{$\dag$}ghzeng@sjtu.edu.cn\\ [1em]
  \footnotesize \textbf{Corresponding author:} Tao Wang\texttt{<tonystar@sjtu.edu.cn>}, Guihua Zeng\texttt{<ghzeng@sjtu.edu.cn>}\\
}

\begin{document}

\twocolumn[ % Method A for two-column formatting
  \begin{@twocolumnfalse} % Method A for two-column formatting

\maketitle
\thispagestyle{empty}

\begin{abstract}
The quantum internet holds the potential to facilitate applications that are fundamentally inaccessible to the classical internet. Among its most prominent applications is quantum key distribution (QKD) networks, which connect two distant nodes to establish a secure key based on the principles of quantum mechanics. However, the subsequent extensive reliance on interferences in existing QKD protocols leads to the weak robustness of the system and the corresponding network. In this work, we propose a robust and cost-effective quantum network using the Kramers-Kronig receiver. We first propose a continuous-variable QKD protocol based on direct detection without interference, which achieves the recovery of quadrature components through the Kramers-Kronig relation. Subsequently, we have extended this protocol to continuous-variable quantum access networks, further highlighting the robustness and cost advantages of interference-free detection. The experimental results show that each user can achieve a secret key rate at $50$ kbit/s within the access network range by using only one photodetector without interference structures. This scheme opens up new possibilities in establishing a robust and cost-effective quantum network, serving as a foundational element in the progress toward establishing a large-scale quantum internet.
\end{abstract}
%\keywords{First keyword \and Second keyword \and More} % (optional)
\vspace{0.35cm}

  \end{@twocolumnfalse} % Method A for two-column formatting
] % Method A for two-column formatting

%\begin{multicols}{2} % Method B for two-column formatting (doesn't play well with line numbers), comment out if using method A

%%%%%%%%%%%%%%%  Main text   %%%%%%%%%%%%%%%
% \linenumbers

\section*{Introduction}

The quantum internet aims to enable capabilities beyond the reach of classical networks, with quantum key distribution (QKD) serving as a foundational pillar for secure connectivity \cite{wootters1982single,hall1995information,kimble2008quantum,wehner2018quantum}. QKD has matured from point-to-point links to scalable networks, broadly categorized into discrete-variable (DV) \cite{scarani2009security,diamanti2016practical,zhang2018large,xu2020secure,lo1999unconditional,lo2014secure} and continuous-variable (CV) \cite{polkinghorne1999continuous,ralph2000security,grosshans2002continuous,grosshans2002reverse,grosshans2003quantum} protocols. Practical implementations, such as the Beijing-Shanghai trunk line \cite{chen2021integrated} and networks in Europe and Japan \cite{stucki2011long,sasaki2011field,dynes2019cambridge}, have demonstrated the viability of these technologies. However, a significant barrier to widespread adoption—particularly in the access layer—is the reliance on interferometric structures for quantum state measurement.

Interference stability is a ubiquitous challenge in existing QKD architectures. Phase-encoding DV-QKD relies on single-photon interference \cite{townsend1994quantum,yuan2005continuous,ribordy1998automated,mo2005faraday,zhou2003time,wang2018practical}, while CV-QKD employs coherent detection that necessitates interference between the signal and a local oscillator (LO) \cite{huang2020experimental,huang2021realizing,wang2023experimental,xu2023round,qi2024experimental,li2024experimental,hajomer2024continuous,pan2025high}. Maintaining stable interference conditions is technically demanding; environmental perturbations such as temperature fluctuations and mechanical vibrations can induce severe phase misalignment or frequency drift. While these issues are manageable in static point-to-point links via complex stabilization mechanisms \cite{jouguet2013experimental,fossier2009field,lodewyck2007quantum,qi2007experimental,huang2015high,huang2016continuous,qi2015generating}, they become exponentially difficult to manage in dynamic, multi-user networks involving routers and switches. This vulnerability fundamentally limits the robustness and cost-effectiveness of large-scale quantum access networks (QANs).

In contrast, classical optical networks predominantly utilize direct detection due to its inherent robustness and simplicity \cite{agrell2016roadmap,zhong2018digital}. While early polarization-based DV-QKD attempted direct detection \cite{bennett2014quantum}, it suffered from fiber birefringence and dispersion. To address the robustness bottleneck in CV-QKD without sacrificing performance, we propose an interference-free architecture inspired by the Kramers-Kronig receiver principles \cite{mecozzi2016kramers,chen2018kramers,harter2020generalized,qu2018continuous,pousset2025kramers}.

In this paper, we present a direct-detection-based CV-QKD (DD CV-QKD) scheme that utilizes commercial photodetectors (PDs) to eliminate the need for interferometric structures. We provide a rigorous security proof by establishing the equivalence between our direct detection operators and standard heterodyne detection. Furthermore, we extend this architecture to a direct-detection-based CV quantum access network (DD CV-QAN). We validate the scheme experimentally and the experimental results show that each user can achieve a secret key rate (SKR) at $50$ kbit/s. Our results demonstrate that DD CV-QAN significantly enhances system robustness and reduces detection costs, offering a scalable solution for future quantum access networks.

\section*{DD CV-QKD}
In conventional CV-QKD protocols, secure key distribution relies on encoding and measuring the quadrature components of the optical field. Typically, the transmitter modulates key information onto the quadrature-$x$ and quadrature-$p$ components, while the receiver employs coherent detection—interfering the signal with a LO—to measure these quadratures.

Here, we propose a direct detection scheme that integrates a Kramers-Kronig receiver into the CV-QKD framework. This approach eliminates the need for physical interferometry (i.e., mixing with an LO). Instead, the scheme reconstructs the complex optical field, including the quadrature components, solely by detecting the optical intensity of the signal. We term this architecture the DD CV-QKD scheme.

\subsection*{Scheme} 
The DD CV-QKD scheme is founded on the Kramers-Kronig relation, which allows the phase of a complex signal to be uniquely derived from its intensity envelope, provided the signal is minimum-phase. The minimum-phase condition requires that the signal's trajectory in the complex plane does not encircle the origin \cite{mecozzi2016kramers}. Detailed procedures for constructing the minimum-phase signal and applying the KK relation are provided in \textbf{supplementary note 1}. The protocol proceeds as follows:

\noindent\rule[3pt]{\columnwidth}{0.1em}

\noindent\textbf{Protocol:} DD CV-QKD

\noindent\rule[3pt]{\columnwidth}{0.05em}

\begin{itemize}[
	leftmargin=3.6em,    
	labelwidth=2.5em,
	labelsep=0.5em,
	topsep=2pt,
	parsep=2pt,
	itemsep=4pt,
	before=\setlength{\parindent}{0em}
	]
	
	\item[\textbf{Step 1:}]
	\textbf{State Preparation.} The transmitter generates two sets of Gaussian random numbers $a(n)$ and $b(n)$ with zero mean. These are combined as $c=a(n)+ib(n)$ and upsampled using a raised-cosine pulse filter to obtain the Gaussian modulated signal with bandwidth $B$. To satisfy the minimum-phase condition, the Gaussian modulated signal and a DC component are loaded onto the optical field with angular frequencies of $\omega_s$ and $\omega_l$, respectively. The resulting field, denoted as $\hat E_{0}(t)$, is transmitted over the quantum channel.
	
	\noindent\makebox[\columnwidth][l]{\hspace*{-3.6em}\rule[3pt]{\columnwidth}{0.05em}}
	
	\item[\textbf{Step 2:}]
	\textbf{Detection.} At the receiver, photoelectric conversion is employed for the direct detection of the optical signal. This process yields the photocurrent operator $\hat{I}(t)$, which is proportional to the instantaneous intensity of the received field.
	
	\noindent\makebox[\columnwidth][l]{\hspace*{-3.6em}\rule[3pt]{\columnwidth}{0.05em}}
	
	\item[\textbf{Step 3:}]
	\textbf{Field Reconstruction.} The receiver applies digital signal processing based on the Kramers-Kronig relation to retrieve the phase information from the intensity information of the minimum-phase signal. This allows for the reconstruction of the operator $\hat E_{MPr}(t)$, from which the light field carrying the pulse signal and its quadrature components are restored. Ultimately, the key information encoded in the quadrature components is obtained.
	
	\noindent\makebox[\columnwidth][l]{\hspace*{-3.6em}\rule[3pt]{\columnwidth}{0.05em}}
	
	\item[\textbf{Step 4:}]
	\textbf{Post-processing.} The transmitter and receiver perform classical post-processing, including reverse reconciliation and privacy amplification, to extract the final secure secret key.
	
	\noindent\makebox[\columnwidth][l]{\hspace*{-3.6em}\rule[3pt]{\columnwidth}{0.05em}}
\end{itemize}

For a detailed mathematical derivation of the modulation, reception, and recovery processes, please refer to \textbf{supplementary note 2}.

\begin{figure}[!t]
	\centering
	\includegraphics[width=\linewidth]{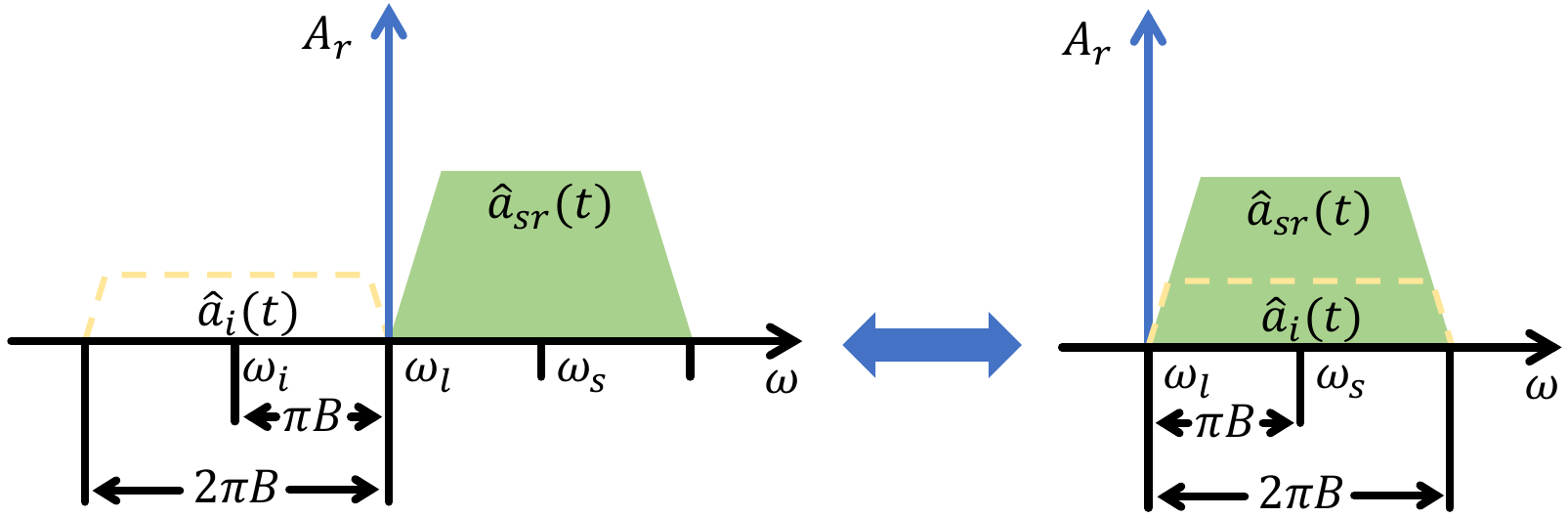}
	\caption{Equivalent representation of the image-band vacuum fluctuation.}
	\label{fig:image}
\end{figure}

\subsection*{Security Analysis}

To rigorously establish the security of the DD CV-QKD scheme, we examine the measurement operators associated with the Kramers-Kronig receiver and demonstrate their equivalence to those of conventional heterodyne detection. Pousset et al. demonstrated the equivalence of the Kramers-Kronig detection operator and the ideal heterodyne operator in the time domain \cite{pousset2025kramers}. Considering the complexity of the time-domain expression of the Kramers-Kronig relation, we referred to their method and derived a similar result in the frequency domain. While direct detection typically measures intensity, the Kramers-Kronig relation allows for the reconstruction of the complex optical field, thereby granting access to the quadrature components.

The core of our analysis lies in the treatment of the image-band vacuum fluctuation. In direct detection, the beating of the signal with itself and the DC component introduces vacuum noise from the image frequency band ($\omega_i = 2\omega_l - \omega_s$), which is coherently superimposed onto the signal band \cite{kikuchi2015fundamentals}, illustrated in figure \ref{fig:image}. By mapping this image-band vacuum to an effective noise source added to the signal, we derive the reconstructed field operator $\hat E_{MPr}(t)$.

Our derivation shows that the quadrature operators obtained via Kramers-Kronig reconstruction, denoted as $\hat x_{KK}$ and $\hat p_{KK}$, satisfy the commutation relation $[\hat x_{KK}, \hat p_{KK}] = 0$. Furthermore, due to the contribution of the image-band vacuum, the inherent shot noise of the system is $2V_{vac}$ (where $V_{vac}$ is the vacuum variance), which is double that of a single quadrature measurement but identical to the total noise in heterodyne detection.

After performing shot-noise normalization, the normalized operators are given by:
\begin{equation}
	\begin{aligned}
		&\hat x_{KK}\leftarrow\frac{\hat x_{KK}}{\sqrt{2V_{vac}}}
		=\frac{\hat x_{a_{sr}(t)}+\hat x_{a_i(t)}}{\sqrt{2V_{vac}}},
		\\&\hat p_{KK}\leftarrow\frac{\hat p_{KK}}{\sqrt{2V_{vac}}}
		=\frac{\hat p_{a_{sr}(t)}+\hat p_{a_i(t)}}{\sqrt{2V_{vac}}}.
	\end{aligned}
\end{equation}
where $\hat x_{a_{sr}(t)}, \hat p_{a_{sr}(t)}$ are the signal quadratures and $\hat x_{a_i(t)}, \hat p_{a_i(t)}$ are the image-band vacuum quadratures. These expressions are formally identical to the normalized operators of ideal heterodyne detection. Consequently, the prepare-and-measure (PM) model of our scheme is equivalent to that of heterodyne CV-QKD, allowing its security to be guaranteed by existing proofs.

For a comprehensive analysis covering both the rigorous mathematical derivation of these operators and the practical countermeasures against specific implementation vulnerabilities (such as DC component attacks and modulator leakage), please refer to \textbf{supplementary note 3}.

\subsection*{Secret Key Rate Calculation}

Given the established equivalence between DD CV-QKD and heterodyne detection, the security analysis directly follows the standard formalism of heterodyne CV-QKD. Practical detector imperfections, including non-unity quantum efficiency and electronic noise, are modeled by coupling the signal with an Einstein-Podolsky-Rosen (EPR) state via a beam splitter (BS), consistent with the standard entropic calibration method (see \textbf{supplementary note 4} for details).

Under the assumption of collective attacks, the asymptotic SKR using reverse reconciliation is given by \cite{laudenbach2018continuous}:
\begin{equation}
	\mathrm{SKR}=f(\beta I_{\mathrm{AB}}-\chi_{\mathrm{BE}}),
	\label{con:SKR}
\end{equation}
where $f$ is the symbol rate, $\beta \in (0,1)$ is the reconciliation efficiency, $I_{\mathrm{AB}}$ is the mutual information between Alice and Bob, and $\chi_{\mathrm{BE}}$ is the Holevo bound on the information accessible to Eve.

To account for finite-size effects \cite{leverrier2010finite,xu2024robust}, the SKR is formulated as:
\begin{equation}
	\mathrm{SKR_{fs}}=f\frac{n}{N}(\beta I_{\mathrm{AB}}-\chi_{\mathrm{BE}}-\Delta(n)),
\end{equation}
where $N$ is the total data block length, $n$ is the effective block length for key generation (excluding parameter estimation), and $\Delta(n)$ is the correction term associated with privacy amplification security. Detailed calculation procedures are provided in \textbf{supplementary note 4}.

\section*{DD CV-QAN}
To demonstrate the scalability and cost-effectiveness of the proposed scheme, we extend the point-to-point framework to a point-to-multipoint scenario: the DD CV-QAN. This architecture connects a central network node, designated as the quantum line terminal (QLT), to multiple end-users, referred to as quantum network units (QNUs). Here, we detail the implementation and security considerations of this network topology.

\subsection*{Scheme}
Our DD CV-QAN operates in a downstream configuration. The QLT functions as the transmitter, modulating and transmitting the quantum signals. These signals are distributed to $N$ independent QNUs via a $1 \times N$ optical splitter. Each QNU acts as a receiver, employing the direct detection module to measure the incoming quantum states.

The protocol execution mirrors the steps of the point-to-point DD CV-QKD scheme, with the primary distinction being the channel model, which now includes the splitting ratio of the network. Following quantum transmission, the QLT performs independent classical post-processing—including reverse reconciliation and privacy amplification—with each QNU individually to establish distinct secure keys. 

\begin{figure*}[!t]
	\centering
	\includegraphics[width=16cm]{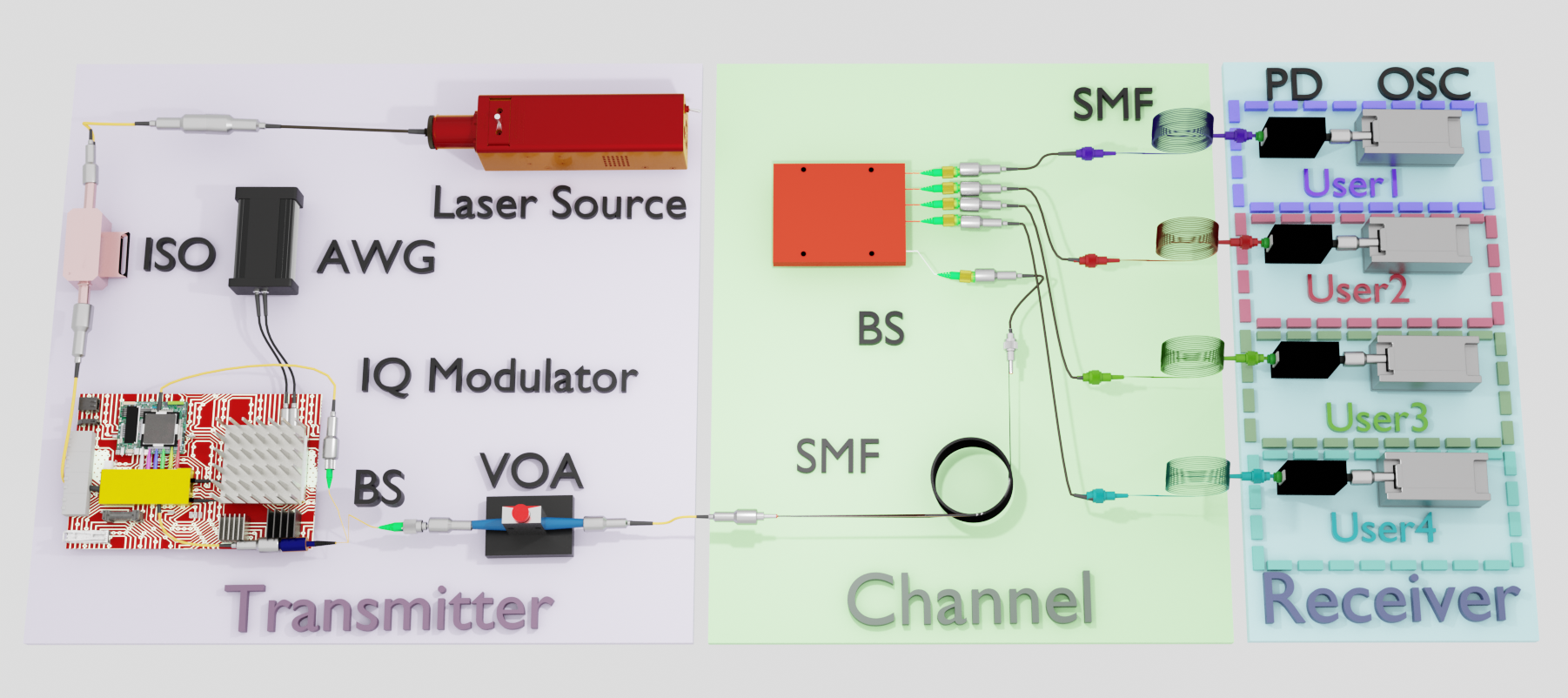}
	\caption{Schematic of the DD CV-QAN system. (ISO: isolator; IQM: IQ modulator; VOA: variable optical attenuator; PD: photodetector; AWG: arbitrary waveform generator; ABC: automatic bias controller; OSC: oscilloscope; SMF: single-mode fiber).}
	\label{fig:optical_pathway_diagram}
\end{figure*}

\subsection*{Security and Performance Analysis}
As the network topology evolves from point-to-point to point-to-multipoint, the security analysis must account for potential information leakage not only to an external eavesdropper (Eve) but also to other users within the network. Following the security framework for downstream CV-QANs established by Pan et al. \cite{pan2025high}, we consider the scenario where the QLT (Alice) generates a key with the $i$-th QNU (Bob$_i$). The system must secure the key against both Eve and other trusted but non-cooperating users (Bob$_j$). The asymptotic SKR is determined by subtracting the maximum information accessible to either party:
\begin{equation}
	\mathrm{SKR}_{AB_i}=f(\beta I_{AB_i}-\max\{\chi_{B_iE},I_{B_iB_j}\}),
\end{equation}
where $I_{AB_i}$ is the mutual information between Alice and Bob$_i$, $\chi_{B_iE}$ is the Holevo bound on Eve's information, and $I_{B_iB_j}$ represents the mutual information between Bob$_i$ and other users. In typical experimental regimes, the external eavesdropping term dominates ($\chi_{B_iE} > I_{B_iB_j}$), simplifying the expression to the conventional form:
\begin{equation}
	\mathrm{SKR}_{AB_i}=f(\beta I_{AB_i}-\chi_{B_iE}).
\end{equation}

While the theoretical SKR formalism for DD CV-QAN is identical to that of heterodyne-based networks, practical factors such as optical splitting losses, electronic noise distribution, and bandwidth limitations introduce specific system differences, which are analyzed in \textbf{supplementary note 5}.

The defining advantage of our DD CV-QAN lies in the superior robustness and cost-effectiveness of the user-side (QNU) hardware. By adopting the Kramers-Kronig receiver, the QNU eliminates the need for interferometric structures, thereby achieving immunity to phase noise and simplifying the optical frontend (detailed robustness analysis in \textbf{supplementary note 6}). Furthermore, replacing complex Balanced Homodyne Detectors (BHDs) or expensive single-photon avalanche diodes (SPADs) with standard commercial photodetectors (PDs) significantly lowers the cost barrier. This design aligns the QNU hardware with the optical network units (ONUs) widely deployed in classical passive optical networks, facilitating seamless integration with existing infrastructure. A detailed cost-effectiveness analysis is provided in \textbf{supplementary note 7}.

\begin{figure*}[!t]
	\centering
	\subfloat[]{\includegraphics[width=5.5cm]{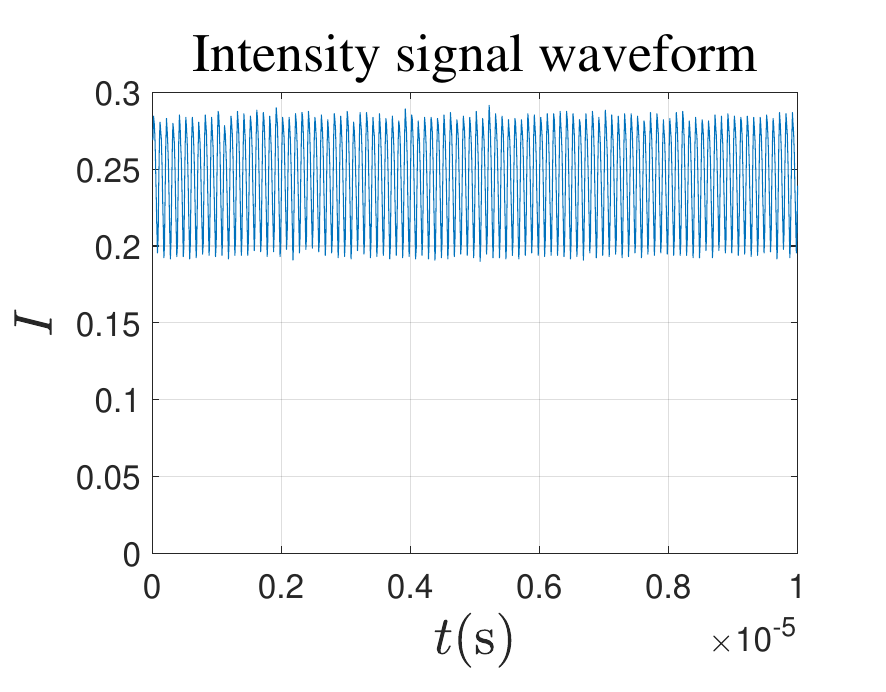}\label{Intensity_signal}}
	\hfill
	\subfloat[]{\includegraphics[width=5.5cm]{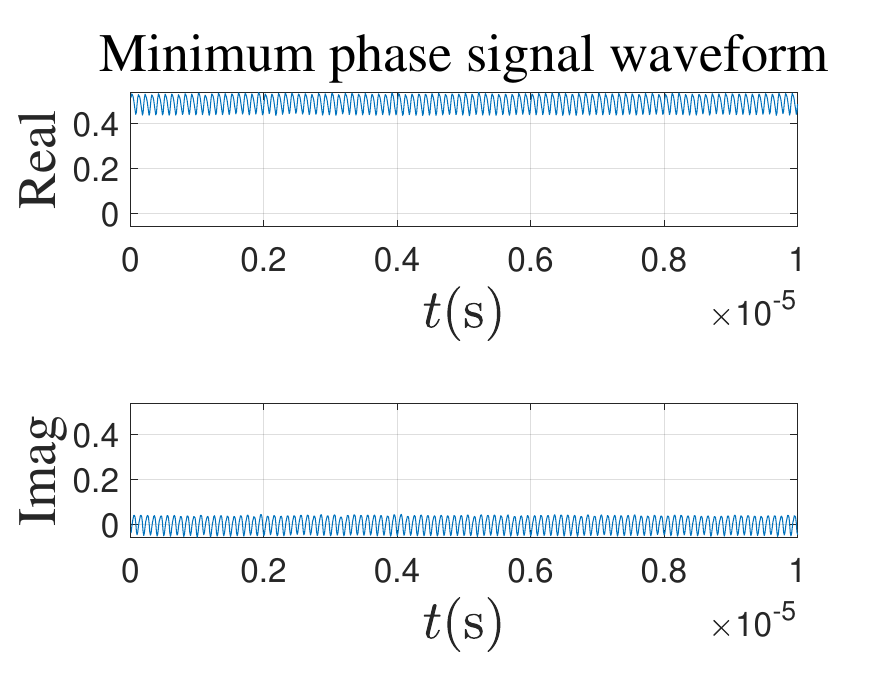}\label{Minimum_phase_signal}}
	\hfill
	\subfloat[]{\includegraphics[width=5.5cm]{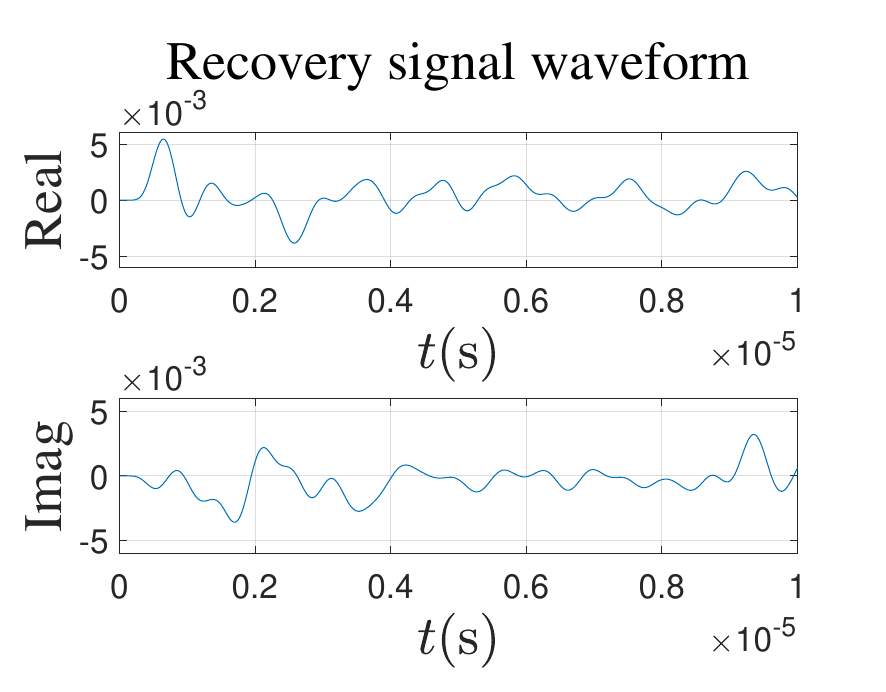}\label{Recovery_signal}}
	
	\centering
	\subfloat[]{\includegraphics[width=5.5cm]{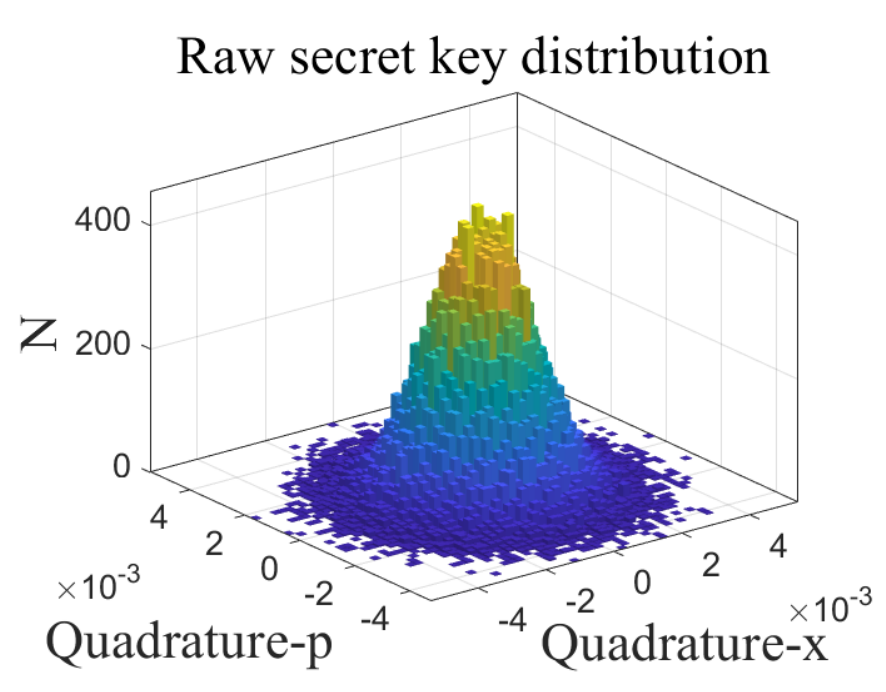}\label{histogram}}
	\hfill
	\subfloat[]{\includegraphics[width=5.5cm]{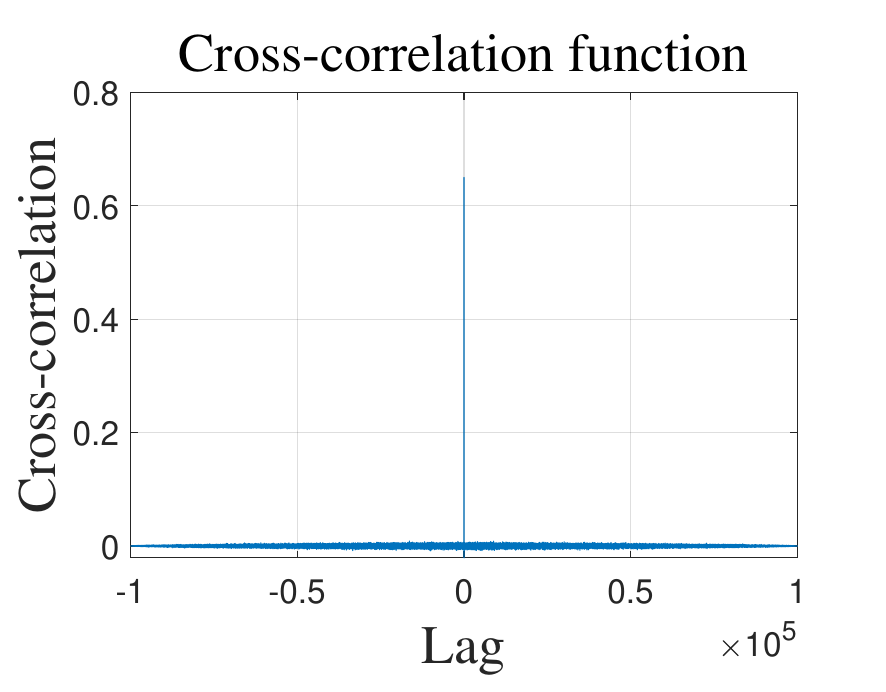}\label{cc}}
	\hfill
	\subfloat[]{\includegraphics[width=5.5cm]{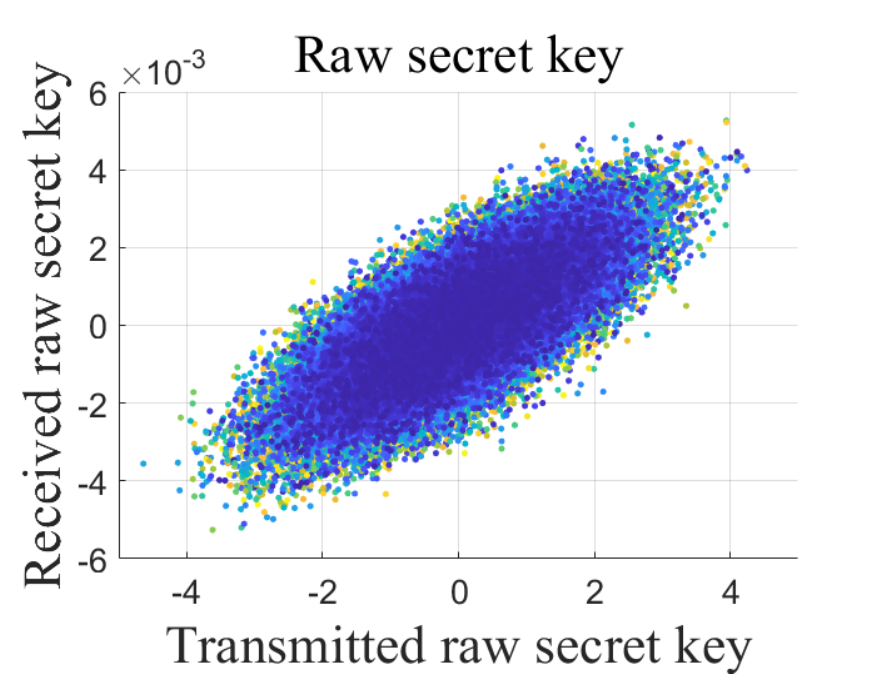}\label{Raw_key}}
	\caption{Data recovery results of DD CV-QAN. Here real and imag denote the real and imaginary parts respectively. (a) The partial time-domain waveform of the intensity signal. (b) The partial time-domain waveform of the minimum-phase signal. (c) The partial time-domain waveform of the recovery signal. (d) The bivariate distribution histogram of the Gaussian data obtained from a frame of received signal. (e) Cross-correlation function between the received data series and the transmitted data series. The cross-correlation function is normalised in such a way that the zero-lag autocorrelation is equal to one. (f) The raw secret key shared by QLT and one of QNUs}
	\label{fig:data_recovery}
\end{figure*}

\begin{figure*}[!t]
	\centering
	\subfloat[]{\includegraphics[width=4.5cm]{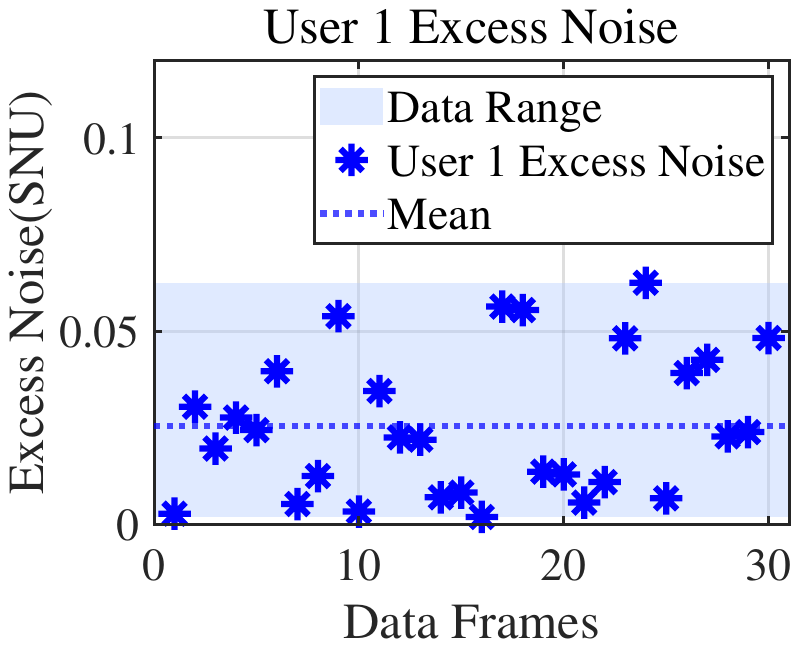}\label{user1}}
	\subfloat[]{\includegraphics[width=4.5cm]{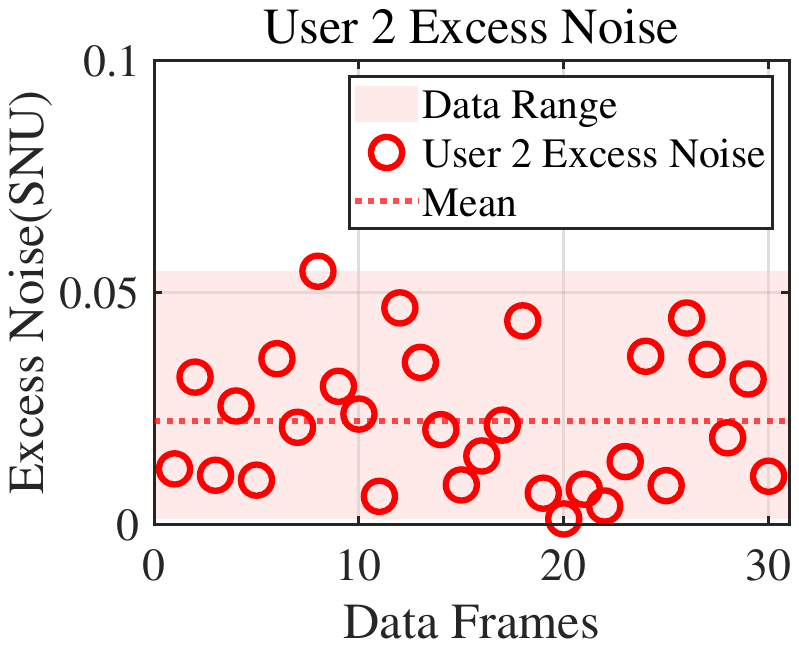}\label{user2}}
	\subfloat[]{\includegraphics[width=4.5cm]{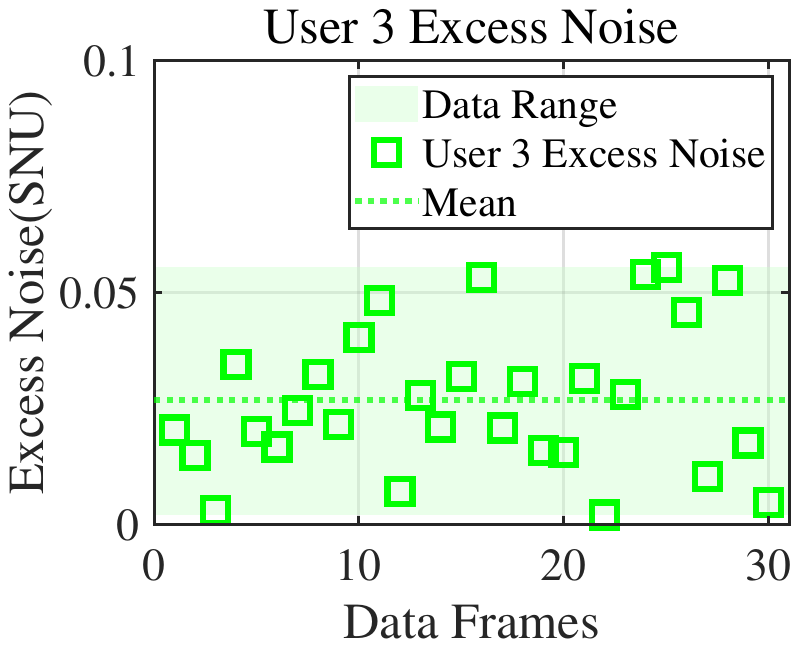}\label{user3}}
	\subfloat[]{\includegraphics[width=4.5cm]{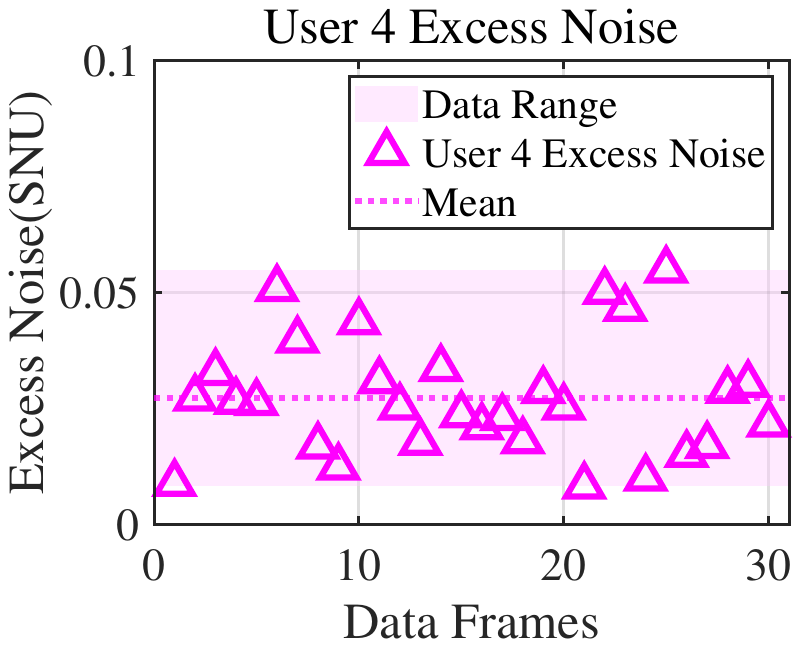}\label{user4}}
	\caption{Excess noise data plots for four users. The horizontal dotted line represents the average excess noise for each user. The shaded areas in each figure represent the data range. (a) The excess noise data for user 1 is represented by blue asterisks. (b) The excess noise data for user 2 is represented by red circles. (c) The excess noise data for user 3 is represented by green squares. (d) The excessive noise data for user 4 is represented by magenta triangles.}
	\label{fig:excess_noise}
\end{figure*}

\section*{Experimental validation}
\subsection*{Experimental set-up}
Figure \ref{fig:optical_pathway_diagram} illustrates the experimental setup of the DD CV-QAN system, comprising three main stages: the transmitter (QLT), the channel, and the receivers (QNUs). 

\textbf{Transmitter (QLT):} A $1550$~nm continuous-wave (CW) laser passes through an isolator (ISO) into an IQ modulator. The modulated output is split by a 90:10 BS; $10\%$ acts as feedback for an automatic bias controller (ABC) to maintain the optimal operating point, while the remaining $90\%$ is attenuated by a Variable Optical Attenuator (VOA) before entering the channel. The driving RF signal is synthesized by an Arbitrary Waveform Generator (AWG, sampling rate $f_s= 100$~MHz). The digital signal processing involves loading random numbers onto a pulse sequence, constructing a minimum-phase signal via spectral shifting and DC component addition, and up-converting it with a carrier frequency $f_{car}=10$~MHz. This carrier facilitates bias control and mitigates leakage from modulator imperfections. The system operates with the symbol rate $f = 1$~MHz, the signal bandwidth $B =10$~MHz, and the DC component $A = 100\sqrt{V_A}$.

\textbf{Channel:} The signal traverses a $2.5$~km feeder fiber to a $1 \times 4$ BS ($6$~dB splitting loss), which distributes the signal equally to four users ($N=4$) via distinct $2.5$~km drop fibers. The single-mode fiber exhibits an attenuation coefficient of $\alpha= 0.21$~dB/km at $1550$~nm.

\textbf{Receiver (QNU):} The receiver architecture implies a cost-effective, robust design requiring only a single photodiode (PD), eliminating the need for a local oscillator or polarization management. Since direct detection measures optical intensity, the system is inherently insensitive to channel-induced phase noise. The Digital Signal Processing (DSP) retrieves the optical field amplitude from the intensity data, reconstructs the phase using the Kramers-Kronig relation, and recovers the original Gaussian signal by removing the DC component and shifting the spectrum to zero frequency. The oscilloscope samples at $f_s=100$~MHz. A data frame comprises $10^5$ symbols. With a PD responsivity of $Re=0.9$~A/W, the quantum efficiency is calculated as $\eta=0.72$ using the relation:
\begin{equation}
	\eta=\frac{hc}{q}\frac{Re}{\lambda},
\end{equation}
where $h$ is Planck's constant, $c$ is the speed of light, and $q$ is the elementary charge.

\subsection*{Data recovery}

We first validate the efficacy of the data recovery process within the DD CV-QAN scheme using a frame of experimental intensity data captured by a user. Figure \ref{fig:data_recovery} illustrates the signal evolution throughout the recovery stages. Specifically, Figure \ref{fig:data_recovery}\subref{Intensity_signal} depicts the time-domain waveform of the detected optical intensity. By applying the Kramers-Kronig relation, the minimum-phase signal is reconstructed (Figure \ref{fig:data_recovery}\subref{Minimum_phase_signal}). Subsequently, the DC component is removed and a spectral shift is applied to obtain the final recovered signal (Figure \ref{fig:data_recovery}\subref{Recovery_signal}), which is then downsampled to generate the raw secret key sequence.

The statistical properties of the recovered keys are analyzed in Figure \ref{fig:data_recovery}\subref{histogram}--\subref{Raw_key}. The histogram in Figure \ref{fig:data_recovery}\subref{histogram} confirms that the raw keys follow a Gaussian distribution in both $x$ and $p$ quadratures. To assess signal fidelity, we calculated the cross-correlation between the transmitted and received key sequences, as shown in Figure \ref{fig:data_recovery}\subref{cc}. A peak correlation coefficient exceeding $0.6$ is observed at zero lag, with negligible values elsewhere, indicating strong synchronization and high randomness of the keys. Furthermore, the scatter plot in Figure \ref{fig:data_recovery}\subref{Raw_key} visualizes the linear relationship between the transmitted and received data, effectively demonstrating the successful reconstruction of the Gaussian modulated signal.

Note that during the reconstruction of the minimum-phase signal, the DC component arriving at the receiver is estimated by averaging the real part of the recovered signal. Further theoretical details are provided in \textbf{supplementary note 8}.

\begin{table}[htbp]
	\centering
	\caption{System parameters used in the experiment.}
	\label{tab:system_parameters}
	
	\small % 字体保持小号
	\renewcommand{\arraystretch}{1.2} % 行高保持

	\begin{tabular*}{\columnwidth}{c @{\extracolsep{\fill}} l l c}
		\toprule
		\textbf{No.} & \multicolumn{2}{c}{\textbf{Parameter}} & \textbf{Value} \\
		\midrule
		1 & \multicolumn{2}{l}{Symbol rate $f$} & 1~MHz \\
		2 & \multicolumn{2}{l}{Modulation variance $V_A$} & 8~SNU \\
		\midrule
		\multirow{4}{*}{3} & \multirow{4}{*}{\ \ \ Attenuation $A_{ch}$} & User 1 & 7.36~dB \\
		& & User 2 & 7.18~dB \\
		& & User 3 & 7.05~dB \\
		& & User 4 & 7.28~dB \\
		\midrule
		\multirow{4}{*}{4} & \multirow{4}{*}{\ \ \ Excess noise $\varepsilon$} & User 1 & 0.0255~SNU \\
		& & User 2 & 0.0223~SNU \\
		& & User 3 & 0.0267~SNU \\
		& & User 4 & 0.0272~SNU \\
		\midrule
		5 & \multicolumn{2}{l}{Quantum efficiency $\eta$} & 0.56 \\
		\midrule
		\multirow{4}{*}{6} & \multirow{4}{*}{\ \ \ Electronic noise $v_{\mathrm{el}}$} & User 1 & 0.0178~SNU \\
		& & User 2 & 0.0180~SNU \\
		& & User 3 & 0.0186~SNU \\
		& & User 4 & 0.0178~SNU \\
		\midrule
		7 & \multicolumn{2}{l}{Reconciliation efficiency $\beta$} & 96\% \\
		\bottomrule
	\end{tabular*}
\end{table}

\subsection*{Parameter evaluation}

Before evaluating the system performance, precise calibration of the shot noise and detector electronic noise is essential.

\subsubsection*{\textbf{Practical shot noise calibration}}
For shot noise calibration, we adopt the methodology established in the theoretical security analysis. By disabling the Gaussian modulation at the transmitter while maintaining the carrier, we effectively measure the vacuum state contributions. To ensure calibration accuracy, the transmitter and receivers are connected via a reference channel with attenuation identical to the experimental link. Further details on the statistical characteristics of the shot noise are provided in \textbf{supplementary note 9}.

\subsubsection*{\textbf{Practical electronic noise calibration}}
Electronic noise calibration follows standard protocols for coherent detection CV-QKD, with specific adaptations for our DSP chain. Since the raw detector electronic noise has a zero mean (which is incompatible with the intensity-based minimum-phase reconstruction), we superimpose a constant bias $C$—corresponding to the average photocurrent of the shot noise calibration—onto the dark noise waveform. This modified signal undergoes the identical DSP pipeline as the actual quantum signal, ensuring equivalent scaling.

We calculated the variances of the two quadrature components from the processed noise data and averaged them to determine the electronic noise power. The calibrated electronic noise levels for the four users are $v_{el1}=0.0178$~SNU, $v_{el2}=0.0180$~SNU, $v_{el3}=0.0186$~SNU, and $v_{el4}=0.0178$~SNU, respectively. Detailed characteristics are discussed in \textbf{supplementary note 10}.

\subsubsection*{\textbf{Modulation variance and excess noise evaluation}}
Following the normalization to SNU, we analyzed the excess noise over 30 data frames, as illustrated in Figure \ref{fig:excess_noise}. Figures \ref{fig:excess_noise}\subref{user1}--\subref{user4} depict the normalized excess noise for each user. The scatter plots illustrate frame-by-frame fluctuations, likely attributable to statistical finite-size effects, while the dotted lines indicate the mean values. The average excess noise levels were measured as $\varepsilon_1=0.0255$~SNU, $\varepsilon_2=0.0223$~SNU, $\varepsilon_3=0.0267$~SNU, and $\varepsilon_4=0.0272$~SNU. Based on these measurements, with a reconciliation efficiency of $\beta=96\%$ and a modulation variance of $V_{A}=8$~SNU, the asymptotic SKR is derived using Eq. (\ref{con:SKR}).

It is worth noting that while our scheme allows for a flexible range of modulation variance (provided the signal remains minimum-phase relative to the DC component), the selected $V_A$ was optimized based on the theoretical models to maximize the SKR.

\begin{table}[htbp]
	\centering
	\caption{Experimental Asymptotic SKR and Simulated Finite-Size SKR Results (kbit/s).}
	\label{tab:skr_results}

	\small
	
	\renewcommand{\arraystretch}{1.2}

	\setlength{\tabcolsep}{0pt}
	
	\begin{tabular*}{\columnwidth}{@{\extracolsep{\fill}} c c c c}
		\toprule
		\multirow{2}{*}{\textbf{User}} & \textbf{Asymptotic} & \multicolumn{2}{c}{\textbf{Finite-Size Simulated}} \\
		\cmidrule(lr){3-4} 
		
		& (Exp.) & $N=10^9$ & $N=10^8$ \\
		\midrule
		User 1 & 49.237 & 22.921 & 19.321 \\
		User 2 & 54.111 & 25.337 & 21.695 \\
		User 3 & 53.357 & 24.992 & 21.407 \\
		User 4 & 49.425 & 23.027 & 19.447 \\
		\midrule
		\textbf{Average} & \textbf{51.533} & \textbf{24.069} & \textbf{20.468} \\
		\bottomrule
	\end{tabular*}
\end{table}

\begin{table*}[t]
	
	\small 
	
	\renewcommand{\arraystretch}{1.2} 
	\centering
	\caption{Comparison of experimental QAN.}
	\label{tab:experimental}
	
	\begin{tabularx}{0.9\textwidth}{>{\centering\arraybackslash}X >{\centering\arraybackslash}X >{\centering\arraybackslash}X >{\centering\arraybackslash}X >{\centering\arraybackslash}X >{\centering\arraybackslash}X >{\centering\arraybackslash}X}
		\hline
		\textbf{Reference} & \textbf{Year} & \textbf{Protocol} & \textbf{Users} & \textbf{Range(km)} & \textbf{SKR(kbit/s)} & \textbf{Detector} \\
		\hline
		\cite{frohlich2013quantum} & 2013 & DV-QKD & 2 & 19.9 & 43.1 & SPAD \\
		\cite{frohlich2015quantum} & 2015 & DV-QKD & 2 & 20 & 33 & SPAD \\
		\cite{wang2021practical} & 2021 & DV-QKD & 3 & 21 & 1.5 & SPAD \\
		\cite{huang2024cost} & 2024 & DV-QKD & 2 & 56 & 71.9 & SPAD \\
		\hline
		\cite{huang2020experimental} & 2020 & CV-QKD & 2 & 12.3 & 22.19 & BHD \\
		\cite{wang2023experimental} & 2023 & CV-QKD & 2 & 25 & 175 & BHD \\
		\cite{xu2023round} & 2023 & CV-QKD & 3 & 30 & 0.82 & BHD \\
		\cite{qi2024experimental} & 2024 & CV-QKD & 4 & 10 & 1010 & BHD \\
		\cite{li2024experimental} & 2024 & CV-QKD & 4 & 15 & 150 & BHD \\
		\cite{hajomer2024continuous} & 2024 & CV-QKD & 8 & 11 & 549.2 & BHD \\
		\cite{pan2025high} & 2025 & CV-QKD & 16 & 6 & 2086 & BHD \\
		\hline
		our work & 2026 & CV-QKD & 4 & 5 & 51.533 & PD \\
		\hline
	\end{tabularx}
\end{table*}

\begin{figure}[!t]
	\centering
	\includegraphics[width=8cm]{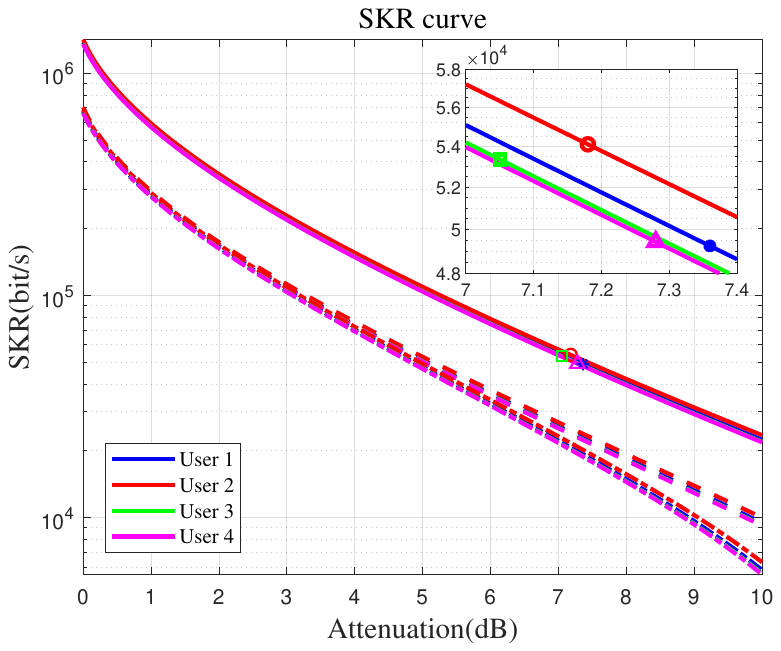}
	\caption{Experimental results of 4-user DD CV-QAN. The blue line, red line, green line, and magenta line represent the SKR curves of user 1, user 2, user 3, and user 4, respectively. The solid line, dashed line, and dash-dot line represent the SKR curves of the asymptotic case, $10^9$ blocks length case, and $10^8$ blocks length case, respectively.}
	\label{fig:skr}
\end{figure}

\subsection*{Experimental SKR Analysis and Comparison}

Table \ref{tab:system_parameters} summarizes the system parameters used to derive the asymptotic SKR curves (solid lines) shown in Figure \ref{fig:skr}. The experimental results for the four users, highlighted in the figure inset, yield asymptotic SKRs ranging from $49.2$ to $54.1$ kbit/s, with an average of $51.5$ kbit/s. To account for finite-size effects due to limited data collection, we simulated block lengths of $10^9$ (dashed lines) and $10^8$ (dash-dot lines). The simulation parameters were set to $n=0.5N$, smoothing parameter $\overline\epsilon=10^{-10}$, and privacy amplification failure probability $\epsilon_{PA}=10^{-10}$. Under these conditions, the SKRs range from $22.9$ to $25.3$ kbit/s for a block length of $10^9$, and decreases to a range of $19.3$ to $21.7$ kbit/s at $10^8$. For specific values, please refer to Table \ref{tab:skr_results}. The relatively restricted SKR is primarily attributed to the $1$~MHz symbol rate. Unlike AC-coupled amplifiers used in standard coherent detection, the DC-coupled amplifiers required here (to preserve the DC component for KK relations) typically impose stricter bandwidth limitations.

Finally, Table \ref{tab:experimental} compares our architecture with existing QAN schemes. DV-QANs rely on single-photon interference and SPADs, which introduce high costs and stability challenges. Similarly, conventional coherent CV-QANs require a local oscillator (LO) and are sensitive to phase drift. In contrast, our DD CV-QAN employs a simplified architecture using a single photodiode (PD) without interferometry. This configuration eliminates the need for phase or polarization locking and significantly reduces costs compared to BHD or SPAD-based systems. Although the current SKR is limited by amplifier bandwidth, the superior robustness and cost-efficiency make this scheme highly advantageous for large-scale access network deployment.

\section*{Discussion}
We have successfully constructed and validated a four-user experimental system based on the DD CV-QAN scheme. Over a total transmission distance of $5$~km with a $1\times4$ splitter, the system achieved an average SKR of approximately $51.5$~kbit/s. 

The primary advantage of the proposed DD CV-QAN lies in the simplified physical layer architecture. Unlike conventional coherent QKD schemes that rely on interferometric structures—which are costly and sensitive to environmental disturbances—our receiver employs a single photodiode for direct detection. This design significantly enhances robustness and reduces deployment costs. Furthermore, since direct detection measures optical intensity, the scheme possesses inherent immunity to phase noise. Consequently, channel-induced phase fluctuations do not affect the detection outcome, eliminating the complex phase tracking or polarization compensation required in standard coherent systems. Although this approach necessitates signal recovery via DSP (specifically the Hilbert transform), the computational load can be efficiently managed by shared processing resources without additional hardware.

However, practical deployment presents specific trade-offs. Firstly, the reconstruction of minimum-phase signals imposes a computational requirement for the Hilbert transform. Secondly, the necessity of measuring the DC component mandates the use of DC-coupled amplifiers. These amplifiers typically offer lower bandwidth compared to the AC-coupled counterparts used in coherent detection, thereby limiting the maximum symbol rate. Nevertheless, unlike backbone optical communications that prioritize ultra-high bit rates, access networks prioritize cost-efficiency and stability, aligning well with the characteristics of our scheme.

In summary, this experiment verifies the feasibility and effectiveness of the DD CV-QAN in an access network environment. While the current analysis simulated finite-size effects to provide tight security bounds, future iterations will focus on implementing higher-speed modulation and collecting larger datasets to experimentally validate finite-size security in real-time.

\section*{Conclusion}
In conclusion, to address the cost and stability challenges associated with interferometric structures in quantum networks, we have proposed a robust and cost-effective DD CV-QKD scheme. By employing a single photodiode combined with the Kramers-Kronig relation, the receiver successfully reconstructs the full complex field information from amplitude measurements, functionally replicating coherent detection without the need for a LO. 

We established the theoretical security of the protocol by analyzing the direct detection operator and demonstrating its equivalence to heterodyne detection. Furthermore, we integrated this protocol into a downstream QAN architecture to maximize receiver-side efficiency. Experimental validation using a four-user system yielded an average SKR of approximately $50$~kbit/s, confirming the feasibility of the principle. 

Ultimately, by eliminating complex interference setups, the proposed DD CV-QAN offers significant advantages in terms of scalability and hardware simplicity. This approach provides a practical and cost-efficient solution for large-scale deployment, paving the way for the future development of the Quantum Internet.

%%%%%%%%%%%%% Acknowledgements %%%%%%%%%%%%%
%\footnotesize
%\section*{Acknowledgements}
%\lipsum[101]

%%%%%%%%%%%%%%   Bibliography   %%%%%%%%%%%%%%

\section*{Acknowledgements}
This work is supported by the National Natural Science Foundation of China [Grant No. 62571316, 61971276], the Quantum Science and Technology-National Science and Technology Major Project (Grant No. 2021ZD0300703), Shanghai Municipal Science and Technology Major Project (Grant No. 2019SHZDZX01), Natural Science Foundation of Shanghai (Grant No. 25ZR1402251),and Cultivation Project of Shanghai Research Center for Quantum Sciences (Grant No. LZPY2024).

\section*{Author contributions}
G.Z. conceived the research; X.L., J.Z., and T.W. performed the theoretical derivation and experimental analysis; X.L., Y.X., and T.W. analyzed the data and wrote the manuscript; Y.X., L.L., and P.H. provided technical support for the experimental platform and data collection. All authors were involved in data collection, discussion of results, and review of the manuscript.

\section*{Competing interests} 
The authors declare no competing interests.

\section*{Additional information}
\textbf{Supplementary Information} is available for this paper at https://doi.org/.
%%%%%%%%%%%%  Supplementary Figures  %%%%%%%%%%%%
% \clearpage

%%%%%%%%%%%% Supplementary Methods %%%%%%%%%%%%
%\footnotesize
%\section*{Supplementary material}
%\lipsum[101]
%
%\lipsum[102]
%
%\lipsum[103]

%%%%%%%%%%%%%%%%   End   %%%%%%%%%%%%%%%%
%\end{multicols}  % Method B for two-column formatting (doesn't play well with line numbers), comment out if using method A
\end{document}

% Use the \preprint command to place your local institutional report
% number in the upper righthand corner of the title page in preprint mode.
% Multiple \preprint commands are allowed.
% Use the 'preprintnumbers' class option to override journal defaults
% to display numbers if necessary
%\preprint{}
%Title of paper
\title{Supplementary Information: Robust and cost-effective quantum network using Kramers-Kronig receiver}
% repeat the \author .. \affiliation  etc. as needed
% \email, \thanks, \homepage, \altaffiliation all apply to the current
% author. Explanatory text should go in the []'s, actual e-mail
% address or url should go in the {}'s for \email and \homepage.
% Please use the appropriate macro foreach each type of information

% \affiliation command applies to all authors since the last
% \affiliation command. The \affiliation command should follow the
% other information
% \affiliation can be followed by \email, \homepage, \thanks as well.
\author{Xu Liu}
\affiliation{State Key Laboratory of Photonics and Communications, Center for Quantum Sensing and Information Processing, Shanghai Jiao Tong University, Shanghai 200240, China}

\author{Tao Wang}
\email{tonystar@sjtu.edu.cn}
\affiliation{State Key Laboratory of Photonics and Communications, Center for Quantum Sensing and Information Processing, Shanghai Jiao Tong University, Shanghai 200240, China}
\affiliation{Shanghai Research Center for Quantum Sciences, Shanghai 201315, China}
\affiliation{Hefei National Laboratory, CAS Center for Excellence in Quantum Information and Quantum Physics, Hefei, Anhui 230026, China}

\author{Junpeng Zhang}
\affiliation{State Key Laboratory of Photonics and Communications, Center for Quantum Sensing and Information Processing, Shanghai Jiao Tong University, Shanghai 200240, China}

\author{Yankai Xu}
\affiliation{State Key Laboratory of Photonics and Communications, Center for Quantum Sensing and Information Processing, Shanghai Jiao Tong University, Shanghai 200240, China}

\author{Yuehan Xu}
\affiliation{State Key Laboratory of Photonics and Communications, Center for Quantum Sensing and Information Processing, Shanghai Jiao Tong University, Shanghai 200240, China}

\author{Lang Li}
\affiliation{State Key Laboratory of Photonics and Communications, Center for Quantum Sensing and Information Processing, Shanghai Jiao Tong University, Shanghai 200240, China}

\author{Peng Huang}
\affiliation{State Key Laboratory of Photonics and Communications, Center for Quantum Sensing and Information Processing, Shanghai Jiao Tong University, Shanghai 200240, China}
\affiliation{Shanghai Research Center for Quantum Sciences, Shanghai 201315, China}
\affiliation{Hefei National Laboratory, CAS Center for Excellence in Quantum Information and Quantum Physics, Hefei, Anhui 230026, China}

\author{Guihua Zeng}
\email{ghzeng@sjtu.edu.cn}
\affiliation{State Key Laboratory of Photonics and Communications, Center for Quantum Sensing and Information Processing, Shanghai Jiao Tong University, Shanghai 200240, China}
\affiliation{Shanghai Research Center for Quantum Sciences, Shanghai 201315, China}
\affiliation{Hefei National Laboratory, CAS Center for Excellence in Quantum Information and Quantum Physics, Hefei, Anhui 230026, China}

%Collaboration name if desired (requires use of superscriptaddress
%option in \documentclass). \noaffiliation is required (may also be
%used with the \author command).
%\collaboration can be followed by \email, \homepage, \thanks as well.
%\collaboration{}
%\noaffiliation

\date{\today}

% insert suggested keywords - APS authors don't need to do this
%\keywords{}

%\maketitle must follow title, authors, abstract, and keywords
\maketitle

% body of paper here - Use proper section commands
% References should be done using the \cite, \ref, and \label commands
%\linenumbers

\section{Supplementary Note 1: Kramers-Kronig relation and minimum-phase signals}

Let the signal $s(t)$ have a bandwidth of $-B/2$ to $B/2$. In order to make this signal satisfy the condition of minimum-phase signal, some transformations need to be applied to it. Remember that the minimum-phase signal obtained from the $s(t)$ transformation is $h(t)$:
\begin{equation}
	h(t)=A+s(t)\exp{(i\pi Bt)},
\end{equation}
where $A$ is a constant. The implication of this operation is that the double-sideband signal $s(t)$ is first spectrally shifted into a single-sideband signal (with a bandwidth of $0$ to $B$), and then a DC component $A$ is added.

In order to recover the phase information from the amplitude information, the relationship between the two needs to be obtained. Write $h(t)$ in magnitude-phase form:
\begin{equation}
	h(t)=|h(t)|\exp{(i\phi_h(t))}.
\end{equation}
Taking the natural logarithm of this yields $u(t)$:
\begin{equation}
	u(t)=\ln[h(t)]=\ln|h(t)|+i\phi_{h}(t).
\end{equation}

According to the proof in the literature \cite{mecozzi2016kramers}, $u(t)$ is a single-sideband signal when the constant $A$ added by constructing $h(t)$ satisfies $|A|>|s(t)|$, and the Hilbert transform relation is satisfied between its real part $ln|h(t)|$ and its imaginary part $\phi_{h}\left(t\right)$:
\begin{equation}
	\phi_{h}(t)=\frac{1}{\pi}p.v.\int_{-\infty}^{+\infty}dt^{\prime}\frac{\ln\left[|h(t^{\prime})|\right]}{t-t^{\prime}},
\end{equation}
where $p.v.$ denotes Cauchy's principal value. This formula is the Kramers-Kronig relation, through which the Kramers-Kronig relation can be obtained from the amplitude $|h(t)|$ of the minimum-phase signal $h(t)$ to its phase $\phi_{h}\left(t\right)$, and finally recover the original complex signal $h(t)$. Then the $h(t)$ is inverted according to its construction process, and finally the original signal $s(t)$ can be obtained.

\begin{figure}
	\centering
	\includegraphics[width=8.5cm]{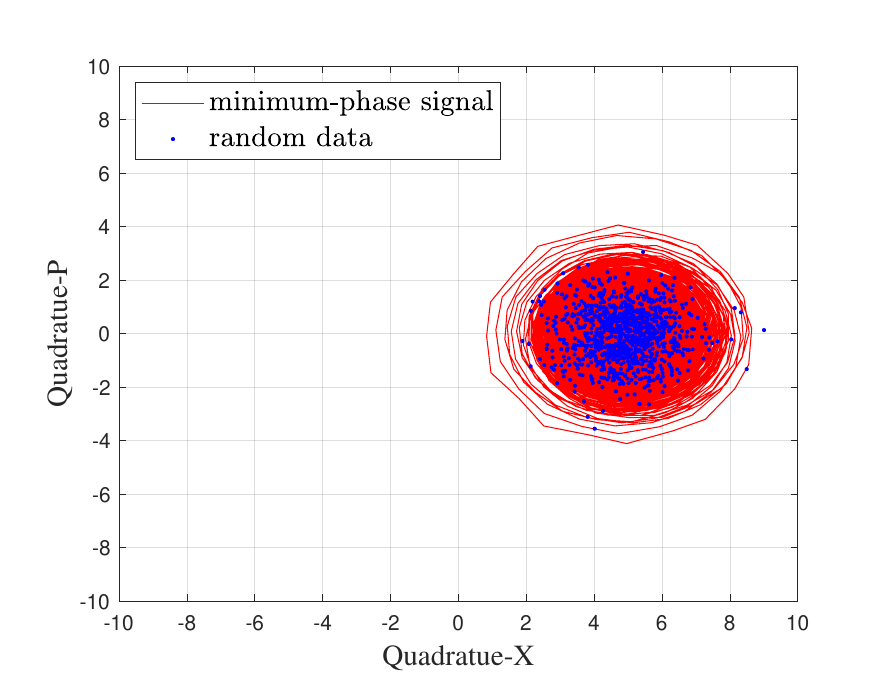}
	\caption{Constellation diagram of Gaussian modulated minimum-phase signals.}
	\label{fig:MPSC}
\end{figure}

We use a set of Gaussian-modulated random numbers as the signal to demonstrate the characteristics of minimum-phase signals. We plot the signal points and the trajectory of the minimum-phase signal in the constellation diagram, where the signal points are shifted by an amount equal to the DC component $A$, as shown in figure \ref{fig:MPSC}. As can be seen in the figure, the red trajectory does not include the origin, which is also an important characteristic of minimum-phase signals. Signals that satisfy this condition can be recovered using the Kramers-Kronig relation.

\section{Supplementary Note 2: Mathematical Principles and Minimum-Phase Signal Construction}

In this note, we provide the detailed mathematical formulation for the modulation, reception, and recovery steps of the DD CV-QKD scheme, followed by a probabilistic analysis of the minimum-phase condition required for the Kramers-Kronig relation.

\subsection{Mathematical Model of the DD CV-QKD Scheme}

Here we elaborate on the specific operators and transformations involved in the three key steps of the scheme.

\textbf{Modulation (Step 1):} The transmitted light field is defined as
\begin{equation}
	\hat E_{0}(t)=\hat E_l(t)+\hat E_s(t)=\hat E_{MP}(t)\exp(i\omega_l t),
\end{equation}
with the components
\begin{equation}
	\hat E_l(t)=A\exp(i\omega_l t), \quad \hat E_s(t)=\hat a_s(t)\exp(i\omega_s t),
\end{equation}
where $\hat a_s(t)$ is a Gaussian modulated quantum signal with bandwidth $B$. The DC component $A$ is a real number chosen to satisfy $|A|^2 \gg \langle \hat a_s^\dagger(t)\hat a_s(t) \rangle$. The frequency relationship is set to $\omega_{IF}=\omega_s-\omega_l\geq \pi B$. In our setup, we take $\omega_{IF}=\pi B$. Extracting the global phase yields the minimum-phase equivalent field $\hat E_{MP}(t)=A+\hat a_s(t)\exp(i\omega_{IF} t)$.

\textbf{Reception (Step 2):} The photocurrent operator obtained after direct detection is
\begin{equation}
	\hat I(t)=\mu\hat E_{0r}^\dagger(t)\hat E_{0r}(t)=\mu\hat E_{MPr}^\dagger(t)\hat E_{MPr}(t),
\end{equation}
where $\mu$ is the photoelectric conversion coefficient. The subscript $r$ denotes the signal after transmission through the channel (e.g., $\hat E_{0r}(t)$ corresponds to $\hat E_{0}(t)$ after channel loss and noise).

\textbf{Recovery (Step 3):} The Kramers-Kronig relation is employed to retrieve the phase information $\hat\phi_{E}(t)$ from the intensity measurements. This is achieved by applying the Hilbert transform:
\begin{equation}
	\begin{aligned}
		\hat\phi_{E}(t)&=\ln\left(\sqrt{\frac{\hat I(t)}{\mu}}\right)*h(t)
		\\&=\ln\left(\sqrt\frac{\hat I(t)}{\mu}\right)*\frac{1}{\pi t}
		\\&=\frac{1}{2\pi}p.v.\int_{-\infty}^{\infty}dt^{\prime}\frac{\ln\left[\frac{\hat I(t’)}{\mu}\right]}{t-t^{\prime}}.
	\end{aligned}
\end{equation}
Using this phase, we reconstruct the field operator:
\begin{equation}
	\hat E_{MPr}(t)=\sqrt{\frac{\hat I(t)}{\mu}}\exp(i\hat\phi_{E}(t)).
\end{equation}
We then remove the carrier and frequency shift to restore the pulsed signal field:
\begin{equation}
	\hat a_{sr}(t)=(\hat E_{MPr}(t)-A_r)\exp(-i\omega_{IF}t).
\end{equation}
Finally, the expectation of the quadrature component operators is obtained as:
\begin{equation}
	\begin{aligned}
		\hat x_{KK}=\frac{1}{2}[\hat a_{sr}(t)+\hat a_{sr}^\dagger(t)],
		\\
		\hat p_{KK}=\frac{1}{2i}[\hat a_{sr}(t)-\hat a_{sr}^\dagger(t)].
	\end{aligned}
\end{equation}
Key information is subsequently extracted from $\hat x_{KK}$ and $\hat p_{KK}$.

\subsection{Construction and Probability Analysis of Minimum-Phase Signals}

The selection of the DC component $A$ is fundamental to the validity of the Kramers-Kronig relation. The condition for a signal to be minimum-phase is that the trajectory of the complex signal does not encircle the origin, which implies that the DC component must be greater than the maximum modulus of the modulation signal \cite{mecozzi2016kramers}.

For Gaussian modulation, the signal amplitude theoretically extends to infinity, meaning an absolute minimum-phase condition cannot be guaranteed with finite power. However, we can ensure the condition is met with high probability. 

The real and imaginary parts of the Gaussian modulated signal points are independent and obey the same distribution:
\begin{equation}
	s=a+ib, \quad a, b \sim \mathcal{N}(0, \sigma^2).
\end{equation}
The amplitude $|s|=\sqrt{a^2+b^2}$ follows a Rayleigh distribution with the probability density function:
\begin{equation}
	f(|s|, \sigma) = \frac{|s|}{\sigma^2} e^{-\frac{|s|^2}{2\sigma^2}}.
\end{equation}
The cumulative distribution function is $F(|s|,\sigma) = 1 - e^{-\frac{|s|^2}{2\sigma^2}}$. For a modulation variance $V_A=\sigma^2$, we set the DC component as:
\begin{equation}
	A=g\sqrt{V_A}.
\end{equation}
The probability $p$ that the signal satisfies the minimum-phase condition ($|s| < A$) is:
\begin{equation}
	p=1-\exp\left(-\frac{g^2}{2}\right).
\end{equation}
When we select $g=5.257$, the success probability is $p=1-10^{-6}$, rendering the probability of failure negligible. In our experiment, we used a conservative value of $g = 100$, which further reduces the probability of construction failure to near zero.

\section{Supplementary Note 3: Security Analysis: Theoretical Derivation and Practical Countermeasures}

In this note, we present a comprehensive security analysis of the DD CV-QKD scheme. We first provide the rigorous mathematical derivation of the detection operators to prove theoretical equivalence with heterodyne detection. Subsequently, we address specific practical security vulnerabilities—including attacks on the DC component and modulator imperfections—and detail the corresponding defensive measures.

\subsection{Theoretical security}

The adoption of direct-detection schemes inevitably modifies the physical observables accessible at the receiver. It is therefore essential to rigorously establish that the information encoded in these observables remains secure. Compared with conventional CV-QKD protocols, the distinguishing feature of our scheme lies solely in the detection mechanism rather than in the state preparation. Accordingly, the security analysis can be naturally carried out by examining the measurement operators associated with Kramers-Kronig direct detection.

Due to the intrinsic complexity of the time-domain formulation of the Kramers-Kronig relation, we conduct our analysis in the frequency domain, which allows for a more transparent and compact derivation. This approach ultimately demonstrates that the resulting measurement operators are equivalent to those of heterodyne detection, thereby guaranteeing the security of the proposed scheme.

%\begin{figure}[!t]
%	\centering
%	\includegraphics[width=\linewidth]{fig/imageband.pdf}
%	\caption{Equivalent representation of the image-band vacuum fluctuation.}
%	\label{fig:image}
%\end{figure}

\subsubsection{\textbf{Derivation of the Kramers--Kronig detection operators}}

Following the signal preparation procedure of our protocol, a Gaussian-modulated quantum signal and a strong DC component are loaded onto optical carriers with angular frequencies $\omega_s$ and $\omega_l$, respectively, forming the input field $\hat E_{0}(t)$. At the receiver, the optical intensity is measured via direct detection. As is well known from classical optical communication theory \cite{kikuchi2015fundamentals}, direct detection introduces an additional vacuum fluctuation associated with the image frequency band, which can be expressed as
\begin{equation}
	\hat E_i(t)=\hat a_i(t)\exp(i\omega_i t),
\end{equation}
where $\hat a_i(t)$ denotes a vacuum-mode annihilation operator. The image-band frequency $\omega_i$ satisfies
$\omega_l-\omega_i=\omega_s-\omega_l=\omega_{IF}$, i.e., $\omega_i=2\omega_l-\omega_s$.

The total input field to the photodetector is therefore given by $\hat E_{0r}(t)+\hat E_i(t)$, yielding the photocurrent operator
\begin{equation}
	\begin{aligned}
		\hat I(t)&=\mu(\hat E_{0r}^\dagger(t)+\hat E_i^\dagger(t))(\hat E_{0r}(t)+\hat E_i(t)) \\
		&\approx\mu[A_r\!\left(\hat a_{sr}(t)+\hat a_{sr}^\dagger(t)+\hat a_i(t)+\hat a_i^\dagger(t)\right)\!\cos\omega_{IF}t \\
		&+A_r^2+\hat a_{sr}^\dagger(t)\hat a_{sr}(t)],
	\end{aligned}
\end{equation}
where terms quadratic in the vacuum operators and the weak signal field have been neglected. This approximation is justified in the regime $A_r^2 \gg \hat a_{sr}^\dagger(t)\hat a_{sr}(t)$ and $A_r^2 \gg \hat a_i^\dagger(t)\hat a_i(t)$, which is satisfied throughout our experimental implementation.

The above expression explicitly shows that the vacuum fluctuation originating from the image band is coherently superimposed onto the detected signal. This observation motivates an equivalent representation in which the image-band vacuum is mapped onto a vacuum contribution at the signal frequency. Specifically, we consider an effective input field
\begin{equation}
	\begin{aligned}
		\hat {E'}_{0r}(t)&=\hat E_{lr}(t)+\hat {E}_{sr}(t)+\hat a_i(t)\exp(i\omega_s t) \\
		&=\hat {E'}_{MPr}(t)\exp(i\omega_l t),
	\end{aligned}
\end{equation}
with
\begin{equation}
	\begin{aligned}
		\hat {E'}_{MPr}(t)&=A_r+(\hat a_{sr}(t)+\hat a_i(t))\exp(i\omega_{IF} t) \\
		&=A_r+\hat {E'}_{sr}(t).
	\end{aligned}
\end{equation}
This equivalence, allows the effect of the image-band vacuum to be treated as an effective vacuum noise directly added to the Gaussian-modulated quantum signal.

The resulting photocurrent operator becomes
\begin{equation}
	\begin{aligned}
		\hat I'(t)&=\mu\hat {E'}_{MPr}^\dagger(t)\hat {E'}_{MPr}(t) \\
		&\approx\mu[A_r(\hat a_{sr}(t)+\hat a_{sr}^\dagger(t)
		+\hat a_i(t)+\hat a_i^\dagger(t))\cos\omega_{IF}t \\
		&+A_r^2+\hat a_{sr}^\dagger(t)\hat a_{sr}(t)
		],
	\end{aligned}
\end{equation}
which is identical to $\hat I(t)$ under the same approximation. Hence, in the following analysis, we use $\hat I(t)$ to denote the detected photocurrent without loss of generality.

Applying the Kramers-Kronig relation amounts to reconstructing the minimum-phase field operator $\hat {E'}_{MPr}(t)$. Since the Kramers-Kronig relation relies on the Hilbert-transform pair between the real and imaginary parts of a single-sideband signal, we first consider
\begin{equation}
	\ln\hat {E'}_{MPr}(t)=\ln|\hat {E'}_{MPr}(t)|+i\hat\phi_{E'}(t).
\end{equation}

Following Ref.~\cite{mecozzi2016kramers}, one can show that $\ln\hat {E'}_{MPr}(t)$ is a single-sideband signal. Expanding it as a Taylor series yields
\begin{equation}
	\ln\hat {E'}_{MPr}(t)
	=\ln A_r+\sum_{n=1}^{\infty}\frac{(-1)^{n+1}}{n}\left(\frac{\hat {E'}_{sr}(t)}{A_r}\right)^n.
\end{equation}
Since the spectrum of $\hat {E'}_{sr}(t)$ vanishes in the negative-frequency domain, all higher-order self-convolutions preserve this property, confirming the single-sideband nature of $\ln\hat {E'}_{MPr}(t)$.

The time domain expression for the Hilbert transform is $h(t)=\frac{1}{\pi t}$, and the corresponding frequency domain expression is $H(\omega)=-i\cdot sgn(\omega)$. Here $sgn(\omega)$ is a symbolic function, taking 1 in the positive frequency domain and -1 in the negative frequency domain. Since the frequency domain expression is more concise and does not require such complex operations as convolution, we consider the proof from the frequency domain perspective. Let $\ln\hat {E'}_{MPr}(t)=\hat u(t)=\hat c(t)+i\hat d(t)$, then its frequency domain is given by the Fourier transform:

\begin{equation}
	\begin{aligned}
		&\hat u(t)\leftrightarrow \hat U(\omega),\ \hat u^\dagger(t)\leftrightarrow \hat U^\dagger(-\omega),
		\\&\hat c(t)\leftrightarrow \hat C(\omega)=\frac{\hat U(\omega)+\hat U^\dagger(-\omega)}{2},
		\\&\hat d(t)\leftrightarrow \hat D(\omega)=\frac{\hat U(\omega)-\hat U^\dagger(-\omega)}{2i}.
	\end{aligned}
\end{equation}
The Hilbert transform of $\hat C(\omega)$ yields
\begin{equation}
	\begin{aligned}
		\hat C(\omega)H(\omega)&=\frac{\hat U(\omega)-\hat U^\dagger(-\omega)}{2i}=\hat D(\omega).
	\end{aligned}
\end{equation}
Then the corresponding transformation in the time domain is
\begin{equation}
	\hat c(t)*h(t)=\hat d(t),
\end{equation}
i.e.
\begin{equation}
	\begin{aligned}
		\hat\phi_{E’}(t)=\frac{1}{2\pi}p.v.\int_{-\infty}^{\infty}dt^{\prime}\frac{\ln\left[\frac{\hat I(t’)}{\mu}\right]}{t-t^{\prime}}.
	\end{aligned}
\end{equation}

We use it to reconstruct the $\hat {E'}_{MPr}(t)$ operator with the photocurrent operator
\begin{equation}
	\hat {E’}_{MPr}(t)=\sqrt{\frac{\hat I(t)}{\mu}}\exp(i\hat\phi_{E’}(t)).
\end{equation}
Further restoration of the light field carrying the pulsed signals gets
\begin{equation}
	\hat a_{sr}(t)+\hat a_i(t)=(\hat {E’}_{MPr}(t)-A_r)\exp(-i\omega_{IF}t).
\end{equation}
The measurement operators for its two quadrature components are
\begin{equation}
	\begin{aligned}
		\hat x_{KK}&=\frac{1}{2}[\hat a_{sr}(t)+\hat a_i(t)+\hat a_{sr}^\dagger(t)+\hat a_i^\dagger(t)]=\hat x_{a_{sr}(t)}+\hat x_{a_i(t)},
		\\\hat p_{KK}&=\frac{1}{2i}[\hat a_{sr}(t)+\hat a_i(t)-\hat a_{sr}^\dagger(t)-\hat a_i^\dagger(t)]=\hat p_{a_{sr}(t)}+\hat p_{a_i(t)}.
	\end{aligned}
\end{equation}
It should be noted that after superposition with the vacuum state, the $\hat x_{KK}$ and $\hat p_{KK}$ operators commute, which is identical to heterodyne detection \cite{kikuchi2015fundamentals}.

\subsubsection{\textbf{Shot-noise normalization}}

During shot-noise calibration, the same detection procedure is applied with the Gaussian modulation switched off. The measured shot-noise units (SNU) are then
\begin{equation}
	\langle\Delta\hat x^2_{KK\text{-}SNU}\rangle
	=\langle\Delta\hat p^2_{KK\text{-}SNU}\rangle
	=2V_{vac}.
\end{equation}
Perform shot noise normalization to obtain
\begin{equation}
	\begin{aligned}
		&\hat x_{KK}\leftarrow\frac{\hat x_{KK}}{\sqrt{2V_{vac}}}
		=\frac{\hat x_{a_{sr}(t)}+\hat x_{a_i(t)}}{\sqrt{2V_{vac}}},
		\\&\hat p_{KK}\leftarrow\frac{\hat p_{KK}}{\sqrt{2V_{vac}}}
		=\frac{\hat p_{a_{sr}(t)}+\hat p_{a_i(t)}}{\sqrt{2V_{vac}}}.
	\end{aligned}
\end{equation}

\subsubsection{\textbf{Comparison with heterodyne detection}}

For comparison, consider conventional heterodyne detection in CV-QKD, whose measurement operators are
\begin{equation}
	\begin{aligned}
		&\hat x_{het}=\frac{1}{\sqrt2}(\hat x_{a_{sr}(t)}+\hat x_{a_i(t)}),\\
		&\hat p_{het}=\frac{1}{\sqrt2}(\hat p_{a_{sr}(t)}+\hat p_{a_i(t)}).
	\end{aligned}
\end{equation}
The SNU when there is no signal input is
\begin{equation}
	\begin{aligned}
		&\langle\Delta\hat x^2_{het-SNU}\rangle=\langle\Delta\hat x^2_{a_i(t)}\rangle=V_{vac},
		\\&\langle\Delta\hat p^2_{het-SNU}\rangle=\langle\Delta\hat p^2_{a_i(t)}\rangle=V_{vac}.
	\end{aligned}
\end{equation}
Perform shot noise normalization to obtain
\begin{equation}
	\begin{aligned}
		&\hat x_{het}\leftarrow\frac{\hat x_{het}}{\sqrt{V_{vac}}}
		=\frac{\hat x_{a_{sr}(t)}+\hat x_{a_i(t)}}{\sqrt{2V_{vac}}},
		\\&\hat p_{het}\leftarrow\frac{\hat p_{het}}{\sqrt{V_{vac}}}
		=\frac{\hat p_{a_{sr}(t)}+\hat p_{a_i(t)}}{\sqrt{2V_{vac}}}.
	\end{aligned}
\end{equation}

Hence, the normalized measurement operators of the direct-detection scheme are formally identical to those of heterodyne detection. As a result, the prepare-and-measure (PM) models of the two schemes are equivalent.

\subsection{Practical security}

We focus on three primary aspects: attacks targeting the DC component, manipulation of the image frequency band, and spectral leakage caused by IQ modulator imperfections.

\subsubsection{Attacks on the DC component}
\begin{figure}[!t]
	\centering
	\includegraphics[width=9cm]{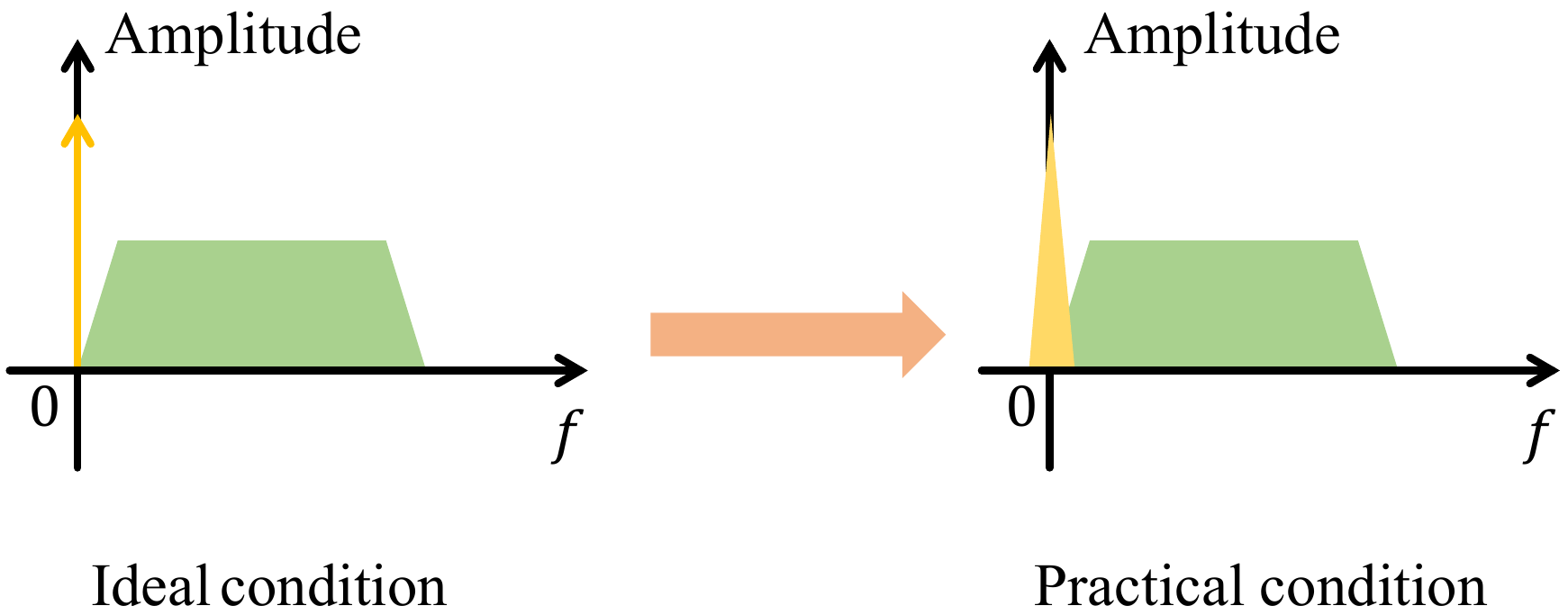}
	\caption{Spectrogram of the minimum-phase signal at the receiver.}
	\label{fig:MPSS}
\end{figure}

In our scheme, the DC component is transmitted through the channel. We consider two attack strategies:

\textbf{1. Intensity manipulation:}
Ideally, the DC component is a delta function, but laser noise broadens it (Fig. \ref{fig:MPSS}). We defend against intensity attacks by monitoring the recovered DC spectral height. Any deviation from the calibration value indicates an attack.

\textbf{2. Additive fluctuations:}
Since the transmitter's inherent laser RIN is calibrated as shot noise, any external fluctuations added by an eavesdropper to the DC component will manifest as excess noise during parameter estimation, thereby revealing the attack.

\subsubsection{Manipulation of the image frequency band}
The image-band vacuum physically propagates through the channel. To prevent Eve from manipulating this mode, a physical filter or waveshaper must be placed before the detector to filter out spectral components in the image band, ensuring the input to the detector at this frequency is a pure vacuum state.

\subsubsection{Imperfections in the IQ modulator}
\begin{figure}[!t]
	\centering
	\subfigure[]{\includegraphics[width=9cm]{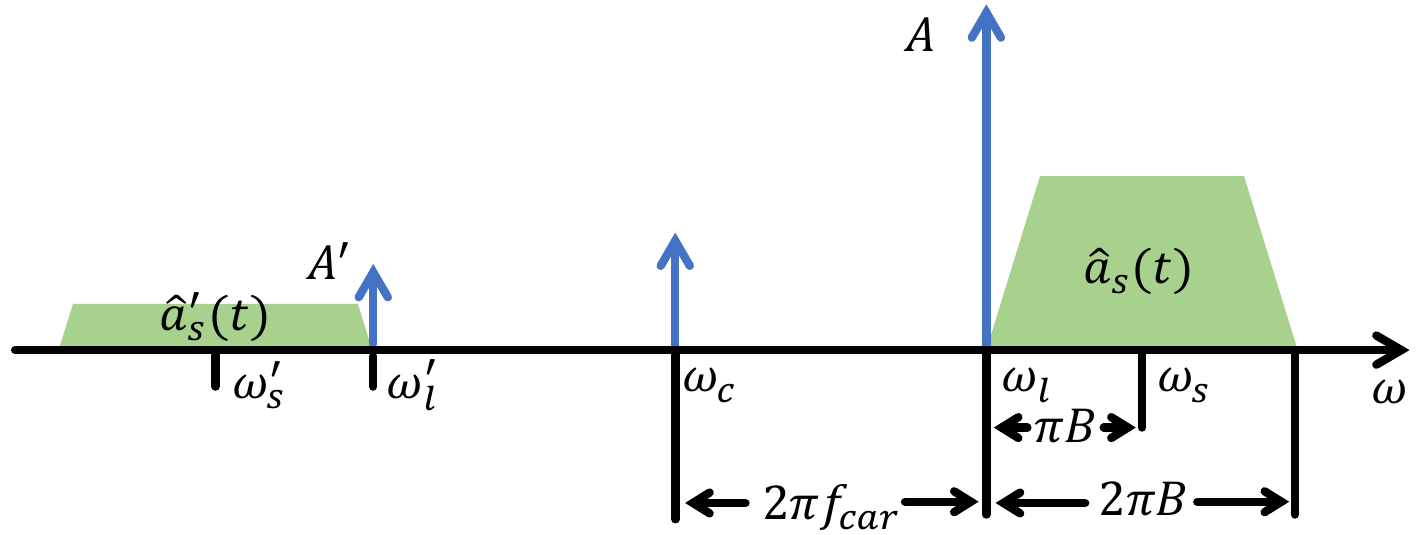}\label{L1}}
	\hfill
	\subfigure[]{\includegraphics[width=9cm]{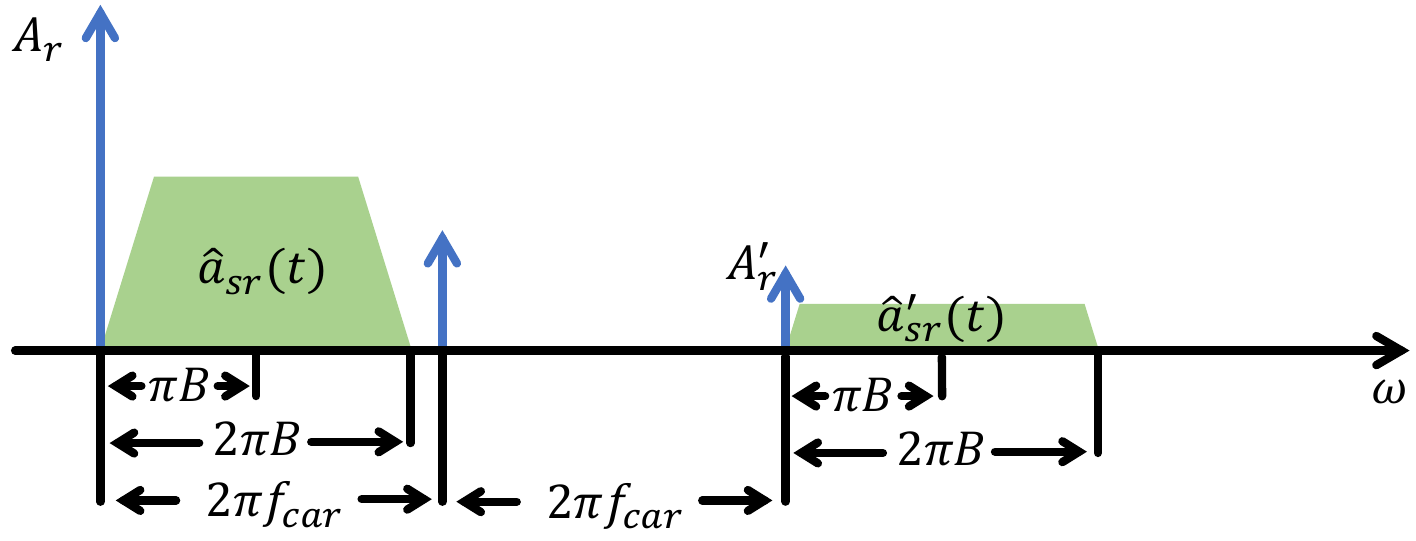}\label{L2}}
	\hfill
	\caption{Spectrum leakage caused by imperfect modulators. (a) Transmitter spectrum after leakage. (b) Equivalent receiver spectrum. }
	\label{fig:leakage}
\end{figure}

Finite extinction ratio in IQ modulators leads to spectral leakage (Fig. \ref{fig:leakage}). This leakage is treated as channel attenuation. By characterizing the leakage ratio and incorporating it into the channel loss assessment, security is maintained even if Eve accesses the leaked sidebands \cite{pereira2018hacking}. Digital or physical filtering at the receiver removes the leakage components from the detection result.

\section{Supplementary Note 4: Secret key rate of homodyne and heterodyne detection}

\begin{figure}[!t]
	\centering
	\subfigure[]{\includegraphics[width=\linewidth]{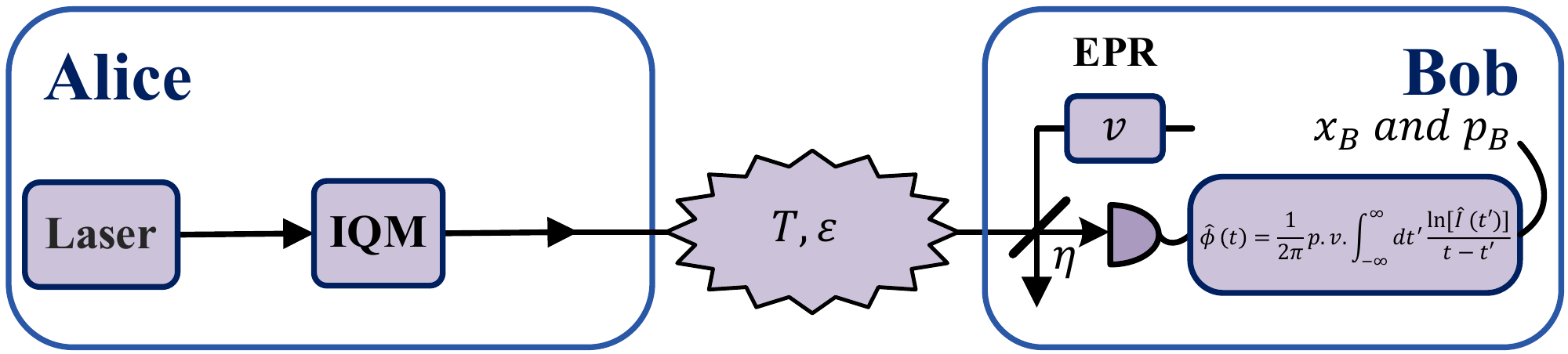}\label{pKKPM}}
	\hfill
	\subfigure[]{\includegraphics[width=\linewidth]{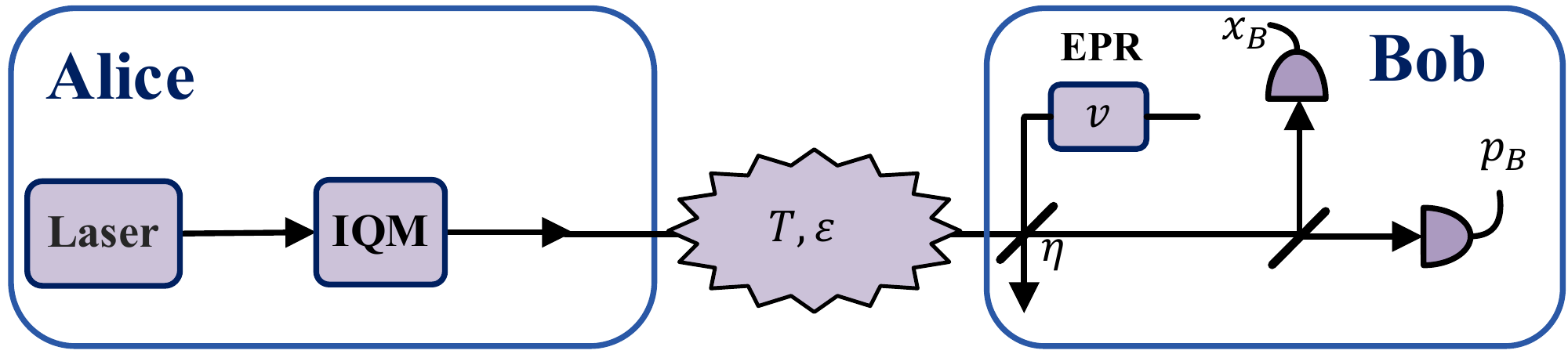}\label{phetPM}}
	\hfill
	\subfigure[]{\includegraphics[width=\linewidth]{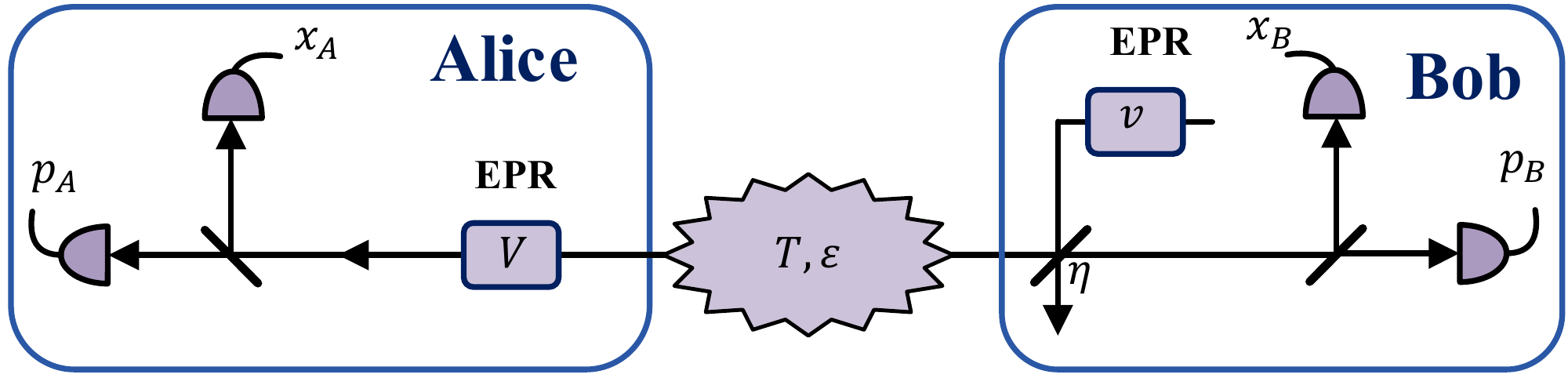}\label{phetEB}}
	\hfill
	
	\caption{CV-QKD model under practical detector conditions. (a) DD CV-QKD prepare-and-measure model. (b) Heterodyne detection CV-QKD prepare-and-measure model. (c) Heterodyne detection CV-QKD entanglement-based model.}
	\label{fig:pmodel}
\end{figure}

In proving security of continuous-variable quantum key distribution (CV-QKD), the entanglement-based (EB) model is often used, which is equivalent to the prepare-and-measure (PM) model used in practical systems. When considering an practical detector with quantum efficiency $\eta$ and electronic noise $v_{el}$, it can be made equivalent by adding a beam splitter (BS) and EPR states to the receiver , as shown in figure \ref{fig:pmodel}. The equivalence relationship still holds here, that is, DD CV-QKD is equivalent to the PM model of heterodyne detection CV-QKD, which in turn is equivalent to the EB model of heterodyne detection CV-QKD.

\begin{figure}
	\centering
	\includegraphics[width=\linewidth]{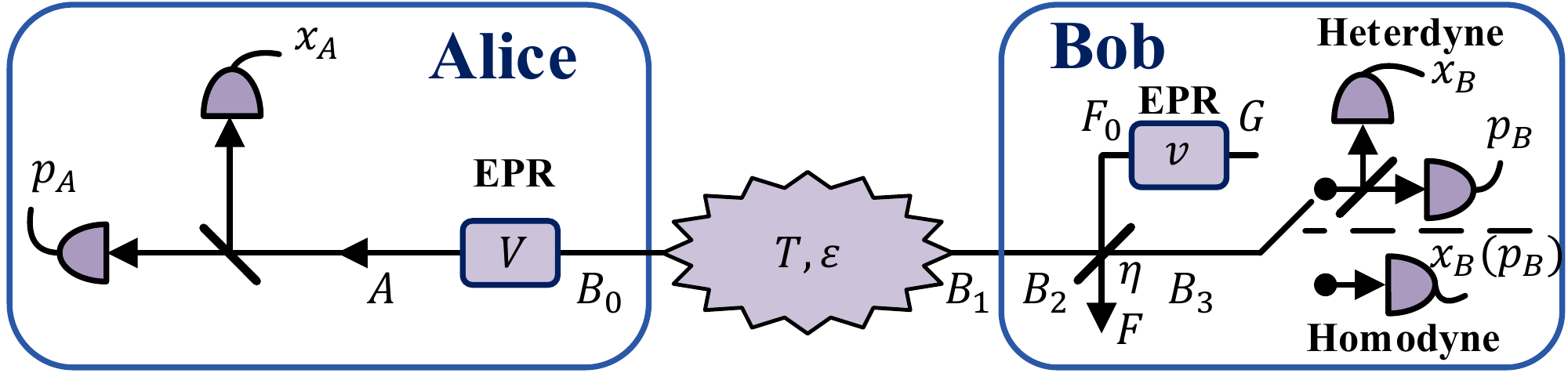}
	\caption{The entanglement-based model of homodyne and heterodyne detection CV-QKD. }\label{fig:EB}
\end{figure}

Here, we introduce the secret key rate (SKR) calculation method for homodyne and heterodyne detection CV-QKD. Our DD CV-QKD scheme uses the same formula as heterodyne detection for calculation \cite{laudenbach2018continuous}. The preparation of a quantum state with modulation variance $V_A$ by Alice at the transmitter is equivalent to heterodyne detection of one of the states $A$ of the two-mode squeezed state with variance $V=V_A+1$, and the other mode $B_0$ is sent to Bob at the receiver.As shown in figure \ref{fig:EB}, the covariance matrix of $AB_0$ is
\begin{equation}
	\gamma_{AB_0}=\left(\begin{array}{cc}\gamma_A&\sigma_{AB_0}\\\sigma_{AB_0}^T&\gamma_{B_0}\end{array}\right)
	=\begin{pmatrix}VI_{2}&\sqrt{T(V^{2}-1)}\sigma_{z}\\\sqrt{T(V^{2}-1)}\sigma_{z}&VI_{2}\end{pmatrix},
\end{equation}
where $I_2=diag(1,1)$, $\sigma_z=diag(1,-1)$. The covariance matrix of $AB_1$ after transmitting through a channel with a transmittance of $T$ and an excess noise of $\varepsilon$ is denoted as $B_1$, and it is given by

\begin{equation}
	\gamma_{AB_1}=\left(\begin{array}{cc}\gamma_A&\sigma_{AB_1}\\\sigma_{AB_1}^T&\gamma_{B_1}\end{array}\right)
	=\begin{pmatrix}VI_{2}&\sqrt{T(V^{2}-1)}\sigma_{z}\\\sqrt{T(V^{2}-1)}\sigma_{z}&T(V+\chi_{line})I_{2}\end{pmatrix},
\end{equation}
where $\chi_{line}=1/T-1+\varepsilon$. At the receiver, the quantum efficiency $\eta$ of the actual detector and the electrical noise are equated to mixing with the two-mode squeezed state $F_0G$ through a BS with a transmittance of $\eta$, and the covariance matrix before mixing is

\begin{equation}
	\gamma_{AB_2F_0G}=\begin{pmatrix}VI_{2}&\sqrt{T(V^{2}-1)}\sigma_{z}&0&0\\\sqrt{T(V^{2}-1)}\sigma_{z}&T(V+\chi_{line})I_{2}&0&0\\0&0&vI_{2}&\sqrt{v^{2}-1}\sigma_{z}\\0&0&\sqrt{v^{2}-1}\sigma_{z}&vI_{2}\end{pmatrix},
\end{equation}
where $v$ is related to the detection noise. For homodyne detection, $\chi_{hom}=[(1-\eta)+v_{el}]/\eta$,$v=\eta\chi_{hom}/(1-\eta)=1+v_{el}/(1-\eta)$. For heterodyne detection, $\chi_{het}=[1+(1-\eta)+2v_{el}]/\eta$,$v=(\eta\chi_{het}-1)/(1-\eta)=1+2v_{el}/(1-\eta)$. The BS is a mixture of $B_1$ and $F_0$ with a matrix of the form

\begin{equation}
	Y_{1}=\begin{pmatrix}I_{2}&0&0&0\\0&\sqrt{\eta}I_{2}&\sqrt{1-\eta}I_{2}&0\\0&-\sqrt{1-\eta}I_{2}&\sqrt{\eta}I_{2}&0\\0&0&0&I_{2}\end{pmatrix}.
\end{equation}
The mixed covariance matrix is

\begin{equation}
	\begin{footnotesize}
		\begin{aligned}
			\gamma_{AB_3FG}&=Y_1\gamma_{AB_2F_0G}(Y_1)^T
			\\&=\begin{pmatrix}VI_{2}&\sqrt{\eta T(V^{2}-1)}\sigma_{z}&-\sqrt{(1-\eta)T(V^{2}-1)}\sigma_{z}&0\\
				\sqrt{\eta T(V^{2}-1)}\sigma_{z}&[\eta T(V+\chi_{line})+(1-\eta)v]I_{2}&\sqrt{\eta(1-\eta)}[v-T(V+\chi_{line})]I_2&\sqrt{(1-\eta)(v^{2}-1)}\sigma_{z}\\
				-\sqrt{(1-\eta)T(V^{2}-1)}\sigma_{z}&\sqrt{\eta(1-\eta)}[v-T(V+\chi_{line})]I_2&[(1-\eta)T(V+\chi_{line})+v\eta]I_{2}&\sqrt{\eta(v^{2}-1)}\sigma_{z}\\
				0&\sqrt{(1-\eta)(v^{2}-1)}\sigma_{z}&\sqrt{\eta(v^{2}-1)}\sigma_{z}&vI_{2}\end{pmatrix}.	
		\end{aligned}
	\end{footnotesize}
\end{equation}
Adjusting its order yield

\begin{equation}
	\begin{footnotesize}
		\begin{aligned}
			\gamma_{AFGB_3}&=\begin{pmatrix}VI_{2}&-\sqrt{(1-\eta)T(V^{2}-1)}\sigma_{z}&0&\sqrt{\eta T(V^{2}-1)}\sigma_{z}\\
				-\sqrt{(1-\eta)T(V^{2}-1)}\sigma_{z}&[(1-\eta)T(V+\chi_{line})+v\eta] I_{2}&\sqrt{\eta(v^{2}-1)}\sigma_{z}&\sqrt{\eta(1-\eta)}[v-T(V+\chi_{line})]I_2\\
				0&\sqrt{\eta(v^{2}-1)}\sigma_{z}&vI_{2}&\sqrt{(1-\eta)(v^{2}-1)}\sigma_{z}\\
				\sqrt{\eta T(V^{2}-1)}\sigma_{z}&\sqrt{\eta(1-\eta)}[v-T(V+\chi_{line})]I_2&\sqrt{(1-\eta)(v^{2}-1)}\sigma_{z}&[\eta T(V+\chi_{line})+(1-\eta)v]I_{2}\end{pmatrix}.
		\end{aligned}
	\end{footnotesize}
\end{equation}
It can be decomposed from this to obtain

\begin{equation}
	\gamma_{AFGB_3}=\left(\begin{array}{cc}\gamma_{AFG}&\sigma_{AFGB_3}\\\sigma_{AFGB_3}^T&\gamma_{B_3}\end{array}\right),
\end{equation}

\begin{equation}
	\gamma_{AFG}=\begin{pmatrix}VI_{2}&-\sqrt{(1-\eta)T(V^{2}-1)}\sigma_{z}&0\\-\sqrt{(1-\eta)T(V^{2}-1)}\sigma_{z}&[(1-\eta)T(V+\chi_{line})+v\eta]I_{2} &\sqrt{\eta(v^{2}-1)}\sigma_{z}\\0&\sqrt{\eta(v^{2}-1)}\sigma_{z}&vI_{2}\end{pmatrix},
\end{equation}

\begin{equation}
	\sigma_{AFGB_3}=\begin{pmatrix}\sqrt{\eta T(V^{2}-1)}\sigma_{z}&\sqrt{\eta(1-\eta)}[v-T(V+\chi_{line})]I_2&\sqrt{(1-\eta)(v^{2}-1)}\sigma_{z}\end{pmatrix}^T.
\end{equation}
If homodyne detection is performed

\begin{equation}
	\gamma_{AFG|B_3}=\gamma_{AFG}-\sigma_{AFGB_3}^T(\Pi_{x}\gamma_{B_3}\Pi_{x})^{MP}\sigma_{AFGB_3},
\end{equation}
where $\Pi_x=\mathrm{diag}(1,0)$ and $^{MP}$ in the expression denotes the Moore-Penrose inverse matrix. If heterodyne detection is performed, first beam splitting of $B_3$ is performed

\begin{equation}
	Y_2=\left(\begin{array}{ccc}{{I_{6}}}&{0}&{0}\\{0}&{{\frac{1}{\sqrt{2}}I_{2}}}&{{\frac{1}{\sqrt{2}}I_{2}}}\\{0}&{{-\frac{1}{\sqrt{2}}I_{2}}}&{{\frac{1}{\sqrt{2}}I_{2}}}\end{array}\right),
\end{equation}

\begin{equation}
	\begin{aligned}
		\gamma_{AFGB_xB_p}&=Y_2\left(\begin{array}{ccc}{{\gamma_{AFG}}}&{{\sigma_{AFGB_3}^T}}&{0}\\{{\sigma_{AFGB_3}}}&{{\gamma_{B_3}}}&{0}\\{0}&{0}&{I_2}\end{array}\right)Y_2^T
		\\&=\begin{pmatrix}\gamma_{AFG}&\frac{1}{\sqrt{2}}\sigma_{AFGB_3}^T&-\frac{1}{\sqrt{2}}\sigma_{AFGB_3}^T
			\\\frac{1}{\sqrt{2}}\sigma_{AFGB_3}&\frac{\gamma_{B_3}+I_2}{2}&\frac{-\gamma_{B_3}+I_2}{2}
			\\-\frac{1}{\sqrt{2}}\sigma_{AFGB_3}&\frac{-\gamma_{B_3}+I_2}{2}&\frac{\gamma_{B_3}+I_2}{2}\end{pmatrix}
		\\&=\begin{pmatrix}\gamma_{AFG}&\sigma_{\alpha}^T&-\sigma_{\alpha}^T
			\\\\\sigma_{\alpha}&\gamma_{\beta}&\sigma_\delta
			\\\\-\sigma_{\alpha}&\sigma_\delta&\gamma_\beta\end{pmatrix}.		
	\end{aligned}
\end{equation}
For $B_x$ measuring the x component and $B_p$ measuring the p component, and since $\gamma_\beta=(\gamma_{B_3}+I_2)/2$ and $\sigma_\alpha=\sigma_{AFGB_3}/\sqrt{2}$, the covariance matrix of $AFG$ becomes:

\begin{equation}
	\begin{aligned}
		\gamma_{AFG|B_x B_p}&=\gamma_{AFG}-\sigma_{\alpha}^T(\Pi_{x}\gamma_{\beta}\Pi_{x})^{MP}\sigma_{\alpha}-\sigma_{\alpha}^T(\Pi_{p}\gamma_{\beta}\Pi_{p})^{MP}\sigma_{\alpha}
		\\&=\gamma_{AFG}-\frac{1}{\gamma_\beta^{11}}\sigma_{\alpha}^T\Pi_x\sigma_{\alpha}-\frac{1}{\gamma_\beta^{22}}\sigma_{\alpha}^T\Pi_p\sigma_{\alpha}
		\\&=\gamma_{AFG}-\sigma_{AFGB_3}^T(\gamma_{B_3}+I_2)^{-1}\sigma_{AFGB_3},
	\end{aligned}
\end{equation}
where $\Pi_{p}=\mathrm{diag}(0,1)$. The total noise from the channel input is $\chi_{tot}=\chi_{line}+\chi_{det}/T$, where $\chi_{det}$ denotes the detection noise, which is taken as $\chi_{hom}$ for homodyne detection and $\chi_{het}$ for heterodyne detection.

Under reverse reconciliation, the secret key rate under asymptotic conditions is

\begin{equation}
	SKR=f(\beta I_{AB}-\chi_{BE}).
	\label{con:SKR}
\end{equation}
where $f$ is the signal repetition frequency. $I_{AB}$ is the mutual information between Alice and Bob, and for homodyne detection, 

\begin{equation}
	I_{AB}^{hom}=\frac{1}{2}\log_2\frac{V+\chi_{tot}}{1+\chi_{tot}}.
\end{equation}
For heterodyne detection,

\begin{equation}
	I_{AB}^{het}=\log_2\frac{V+\chi_{tot}}{1+\chi_{tot}}.
\end{equation}
The $\chi_{BE}$ denotes the maximum amount of information that Eve, the eavesdropper, can obtain from Bob's key:

\begin{equation}
	\chi_{BE}=\sum_{i=1}^2G(\frac{\lambda_i-1}{2})+\sum_{i=3}^5G(\frac{\lambda_i-1}{2}),
\end{equation}
where $G(x)=(x+1)\log_2(x+1)-x\log_2x$, and $\lambda_i$ denotes the symplectic eigenvalue of the covariance matrix. $\lambda_{1,2}$ is the symplectic eigenvalue of $\gamma_{AB_1}$.

\begin{equation}
	\lambda_{1,2}^{2}=\frac{1}{2}(A\pm\sqrt{A^{2}-4B}),
\end{equation}

\begin{equation}
	A=V^{2}(1-2T)+2T+T^{2}(V+\chi_{line})^{2},
\end{equation}

\begin{equation}
	B=T^{2}(V\chi_{line}+1)^{2}.
\end{equation}
$\lambda_{3,4,5}$ is the symplectic eigenvalue of $\gamma_{AFG}^{m_B}$. For homodyne detection $\gamma_{AFG}^{m_B}=\gamma_{AFG|B_3}$, and for heterodyne detection $\gamma_{AFG}^{m_B}=\gamma_{AFG|B_x B_p}$. The symplectic eigenvalue can be calculated by the following equation:

\begin{equation}
	\lambda_{3,4}^{2}=\frac{1}{2}(C\pm\sqrt{C^{2}-4D}),
\end{equation}

\begin{equation}
	\lambda_5=1.
\end{equation}
For homodyne detection, 

\begin{equation}
	C_{hom}=\frac{V\sqrt{B}+T(V+\chi_{line})+A\chi_{hom}}{T(V+\chi_{tot})},
\end{equation}

\begin{equation}
	D_{hom}=\sqrt{B}\frac{V+\sqrt{B}\chi_{hom}}{T(V+\chi_{tot})}.
\end{equation}
For heterodyne detection, 

\begin{equation}
	C_{het}=\frac{A\chi_{het}^2+B+1+2\chi_{het}(V\sqrt{B}+T(V+\chi_{line}))+2T(V^2-1)}{\left(T(V+\chi_{tot})\right)^2},
\end{equation}

\begin{equation}
	D_{het}=\left(\frac{V+\sqrt{B}\chi_{het}}{T(V+\chi_{tot})}\right)^2.
\end{equation}

When finite size effects are considered, the equation \ref{con:SKR} is rewritten as
\begin{equation}
	\mathrm{SKR_{fs}}=f\frac{n}{N}(\beta I_{\mathrm{AB}}-\chi_{\mathrm{BE}}-\Delta(n)),
\end{equation}
where $N$ is the effective data block length, $n$ is the actual data length used to generate the key after removing the parameter estimation part, and $\Delta(n)$ is related to the security of privacy amplification. $\Delta(n)$ can be ignored for the asymptotic case with $\Delta(n)=0$. However, it cannot be neglected for the finite-size case and can be calculated as
\begin{equation}
	\Delta(n)=7\sqrt{\frac{\log_2(1/\overline\epsilon)}{n}}+\frac{2}{n}log_2\frac{1}{\epsilon_{PA}}.
\end{equation}
in which $\overline\epsilon$ denotes a smoothing parameter, and $\epsilon_{PA}$ represents the failure probability of the privacy amplification procedure. 

In addition, under the influence of limited size, it is necessary to correct the transmittance $T$ and excess noise $\varepsilon$ to $T_{min}$ and $\varepsilon_{max}$:
\begin{equation}
	T_{min}=\frac{(\sqrt{\eta T}-\Delta T)^2}{\eta},
\end{equation}
\begin{equation}
	\varepsilon_{max}=\frac{\sigma^2+\Delta\sigma^2-1-v_{el}}{\eta T},
\end{equation}
where $\sigma^2=\eta T+1+v_{el}$. Two offsets due to the finite-size effect are represented as
\begin{equation}
	\Delta T=z_{\epsilon_{PE}/2}\sqrt{\frac{\sigma^2}{mV_A}},
\end{equation}
\begin{equation}
	\Delta \sigma^2=z_{\epsilon_{PE}/2}\frac{\sigma^2\sqrt2}{\sqrt m},
\end{equation}
where $m=N-n$ is the length of the data used for parameter estimation, $z_{\epsilon_{PE}/2}$ is the confidence coefficient.

\section{Supplementary Note 5: Comparison of Kramers-Kronig detection and coherent detection}
According to our security proof, the detection operator of our scheme is equivalent to that of heterodyne detection CV-QKD. However, there are still some differences between the two detection methods in actual systems. Below, we analyze these differences from four aspects: optical losses, electronic noise, electronic bandwidth, and imperfect balancing.

\textbf{1.Optical losses}:

Regarding detection optical loss, the detection loss in Kramers-Kronig detection, similar to coherent detection, is quantified by the quantum efficiency. In the actual experimental process, Kramers-Kronig detection does not need a BS to split the light, and directly changes the optical signal into an electronic signal through photoelectric conversion, and its quantum efficiency is determined by the detector responsivity. The quantum efficiency of coherent detection, on the other hand, tends to be a bit lower because it is additionally affected by the BS. 

\begin{figure}[!t]
	\centering
	\includegraphics[width=9cm]{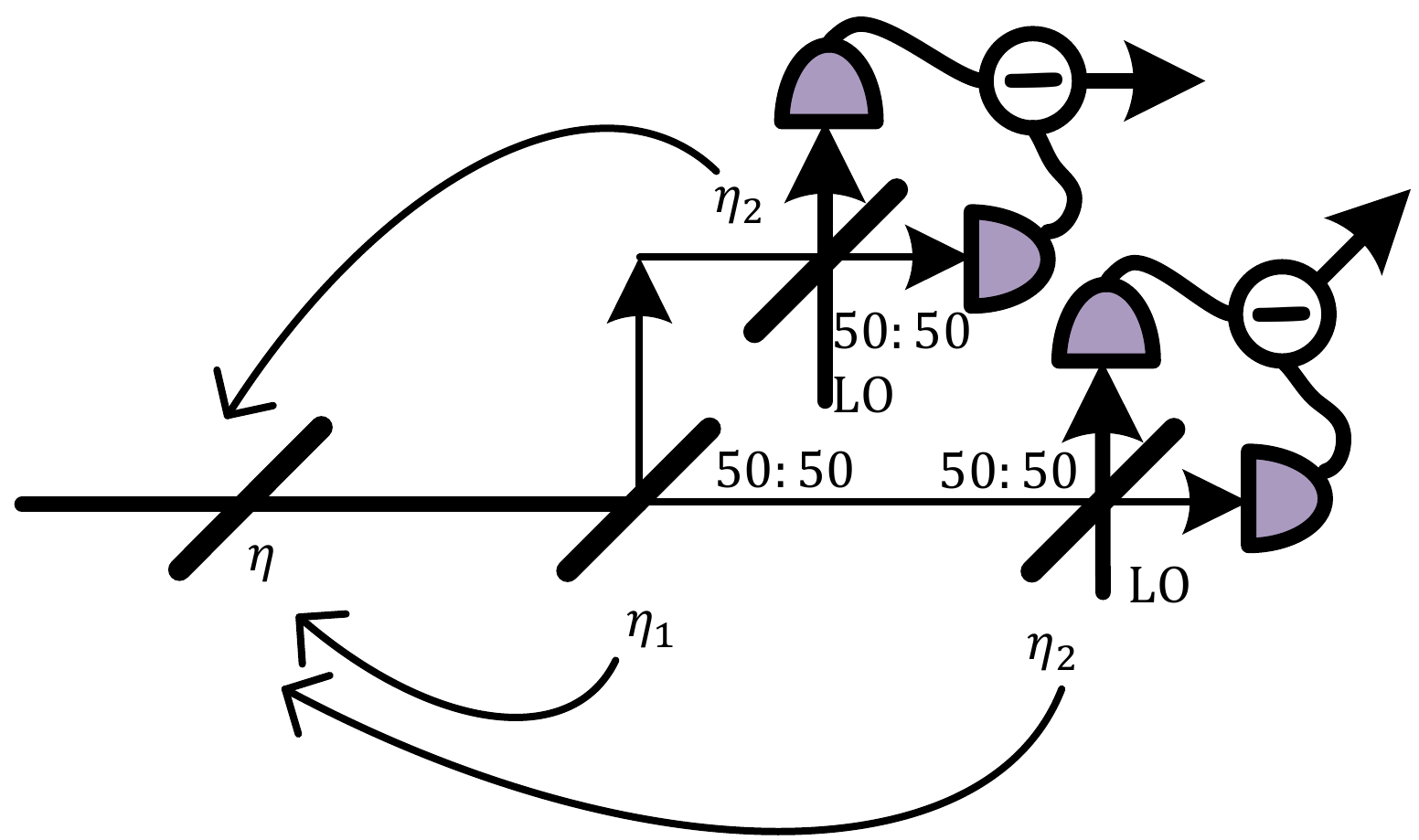}
	\caption{Coherent detector structure.}
	\label{fig:hetdetector}
\end{figure}

Compared with Kramers-Kronig detection, coherent detection requires multiple additional BSs to achieve separate measurements of the two quadrature components and balanced homodyne detection of a single quadrature component, as shown in figure \ref{fig:hetdetector}. Each BS also introduces an additional insertion loss when accessed, as shown by $\eta_1$ and $\eta_2$ in figure \ref{fig:hetdetector}. Here we ignore the imbalance of the detectors and consider the BSs at the end of the measurement of the two quadrature components to be the same. All insertion losses of BS will be reflected in the overall quantum efficiency $\eta$ of the detector, as indicated by the arrows in the figure \ref{fig:hetdetector}. Assuming an insertion loss of $0.5$ dB for each BS, if the photodetectors have the same responsivity, the detection loss of coherent detection will be $1$ dB higher, i.e., the quantum efficiency is about $0.8$ times that of Kramers-Kronig detection. 

\textbf{2.Electronic noise}:

Concerning detector electronic noise, from a physical viewpoint, the source of electronic noise is the noise current of the photodetector and the amplification noise of the amplifier, which are jointly determined by both factors.

For the two schemes of coherent detection and Kramers-Kronig detection, in coherent detection, the quadrature components of the optical field are measured directly, and the electronic noise is superimposed onto these quadrature components independently of the signal. However, for Kramers-Kronig detection, the photocurrent signal obtained by the detector needs to go through a series of operations to obtain the quadrature components, and therefore the electronic noise needs to go through the same computational process.

Furthermore, the processing method for calibrating electronic noise is the same as that used in other coherent detection CV-QKD schemes. Since the signal undergoes a series of digital signal processing steps, electronic noise and shot noise are also processed in the same way during calibration to ensure equivalent scaling with the signal. Since the detector electronic noise is a signal with a mean value of 0, it is not a minimum-phase signal. To measure its contribution here, we add the average value of the photocurrent signal when only the DC component is present, i.e., when calibrating the shot noise, and then perform the same operation. This eliminates the influence of shot noise, allowing us to assess the influence of electronic noise.

In our experiment, we superimposed a constant $C$ on the electronic noise waveform without light input and performed the same digital signal processing as for the signal to obtain data points generated by electronic noise fluctuations. We then calculated the variance of the two quadrature components and took their average to obtain the calibration value of the electronic noise. Here, the constant $C$ is the average value of the shot noise waveform detected in the experiment.

\textbf{3.Electronic bandwidth}:

With regard to the issue of electronic bandwidth, the main focus here is on the analysis of the characteristics of the actual detectors used in the two schemes. For coherent detectors the electronic bandwidth can often reach the GHz level. For Kramers-Kronig detection, it is necessary to measure the DC component in the signal. However, photodetectors capable of detecting DC components cannot achieve the electronic bandwidth level of coherent detectors, and their frequency range is typically between tens of MHz and hundreds of MHz, which limits the repetition frequency of our experimental implementation.

\begin{figure}[!t]
	\centering
	\subfigure[]{\includegraphics[width=8cm]{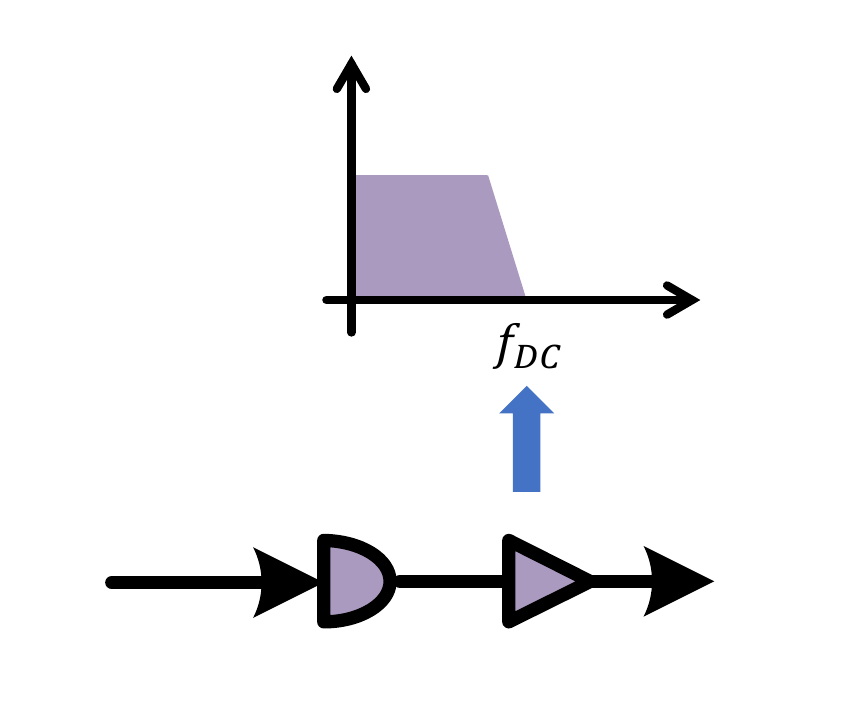}\label{KK}}
	\hfill
	\subfigure[]{\includegraphics[width=8cm]{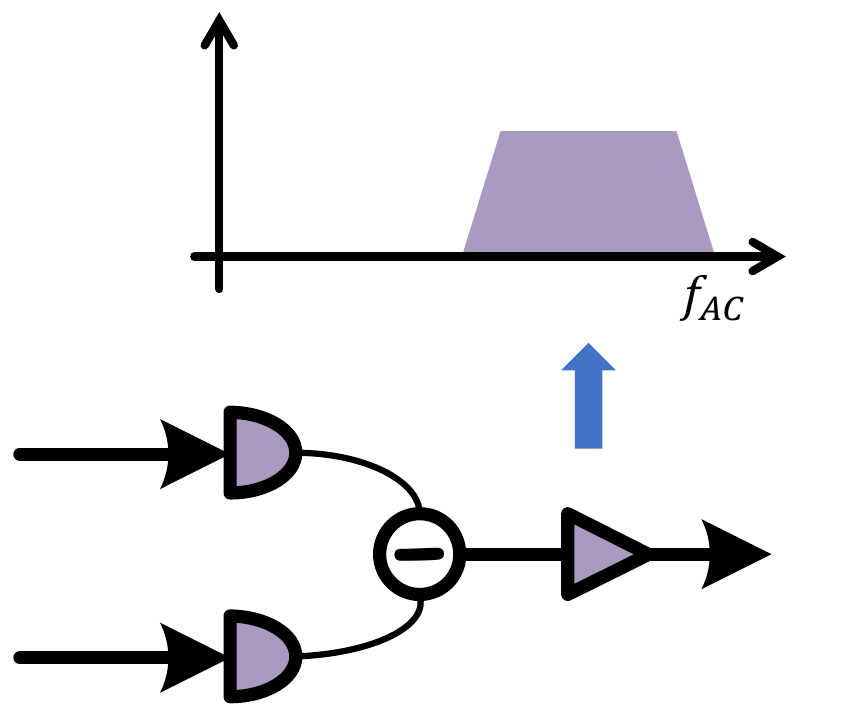}\label{het}}
	\hfill
	\caption{Structure and amplifier bandwidth of two detectors. (a) Photodetectors used in Kramers-Kronig detection and its amplifier bandwidth. (b) Balanced homodyne detector used in coherent detection and its amplifier bandwidth.}
	\label{fig:power_amplifier}
\end{figure}

We further explain the source of the bandwidth difference between the two detectors. Figure \ref{fig:power_amplifier} shows the specific structures of the two detectors and the spectral responses of their amplifiers. Figure \ref{fig:power_amplifier} \subref{KK} shows the photodetector structure used for Kramers-Kronig detection, in which the amplifier must be a DC-coupled amplifier, i.e., the gain bandwidth must include the DC component at zero frequency, which is the main factor limiting its bandwidth. Figure \ref{fig:power_amplifier} \subref{het} shows the balanced homodyne detector structure used for coherent detection, in which the amplifier can be an AC amplifier with a higher bandwidth because coherent detection does not require the measurement of the DC component as in Kramers-Kronig detection.

\textbf{4.Imperfect balancing}:

Regarding the issue of imperfect balance, this occurs when using actual detectors. For coherent detection, a balanced homodyne detector (BHD) consisting of two photodetectors is used, and imbalance between the two detectors introduces additional noise. In addition, in order to measure two quadrature components simultaneously, two BHDs are required for separate detection, and imbalance between the BHDs can also cause different scaling of the measurement results of the two quadrature components, affecting the measurement results.

For Kramers-Kronig detection, although we have proven that it is equivalent to coherent detection of two quadrature components measured simultaneously, in actual implementation, its physical layer structure contains only one photodetector, thus avoiding the problems caused by imperfect balancing in coherent detection.

\section{Supplementary Note 6: QAN Robustness analysis}

We will present a detailed analysis of the robustness associated with our DD CV-QAN scheme. The robustness of the DD CV-QAN scheme is based on the fact that no interference is required in the detection structures in the DD CV-QKD scheme that it uses. We will further illustrate the robustness advantages of our scheme by comparing it with the phase-coded DV-QKD, TLO CV-QKD and LLO CV-QKD, which are widely used in fiber optic systems currently.

For the phase-encoding DV-QKD scheme, two AMZIs are initially used for realization. By combining the two MZI short and long arms with each other, the two optical fields passing through the primary long arm and the primary short arm, respectively, will undergo single-photon interference, and different response results are obtained according to the phase difference of the modulation \cite{townsend1994quantum,yuan2005continuous}. However, the structure is very unstable in practical operation. The arm length of the AMZI will change under the influence of the external environment (e.g., temperature, vibration, etc.), which will lead to the random drift of the phase difference between the two optical fields. The phase difference between the two arms of the AMZI of Alice at the transmitter is $\phi_\mathrm{A}$, and the phase difference between the AMZI of Bob at the receiver is $\phi_\mathrm{B}$. The changes in phase difference generated by the influence of temperature and other external environments are $\Delta\phi_\mathrm{A}$ and $\Delta\phi_\mathrm{B}$, respectively. Here $\Delta\phi_\mathrm{A}+\phi_\mathrm{B}=\Delta\phi_\mathrm{AB}$, where ideally $\Delta\phi_\mathrm{AB}=0$, but the two AMZIs in the physical system can not be the same. Then the phase error of the two optical fields undergoing single-photon interference in addition to the modulation phase difference can be expressed as
\begin{equation}
	\Delta\phi_\mathrm{DV}=\phi_\mathrm{A}+\Delta\phi_\mathrm{A}+\phi_\mathrm{B}+\Delta\phi_\mathrm{B}=\Delta\phi_\mathrm{A}+\Delta\phi_\mathrm{B}+\Delta\phi_\mathrm{AB}.
\end{equation}
Phase drift can lead to bit errors, which in turn affect the robustness of the system. Many improvement schemes have been proposed to address these problems, such as plug and play \cite{ribordy1998automated}, Faraday-Michelson interferometer \cite{mo2005faraday}, Sagnac interferometer \cite{zhou2003time,wang2018practical}, etc. These schemes ensure the stable operation of the point-to-point system.

For TLO CV-QKD \cite{jouguet2013experimental,fossier2009field,lodewyck2007quantum,qi2007experimental}, whether it is time division multiplexing or polarization multiplexing, similar to phase-encoded DV-QKD, it is also a system composed of two AMZIs at the transmitter and receiver. Here, AMZI is used to multiplex signals and LO at the transmitter end and demultiplex them at the receiver end, such as delay lines or polarization beam splitters. The phase error has a similar form:
\begin{equation}
	\Delta\phi_\mathrm{TLO}=\Delta\phi_\mathrm{AT}+\Delta\phi_\mathrm{BT}+\Delta\phi_\mathrm{ABT}.
\end{equation}
Among them, $\Delta\phi_\mathrm{AT}$ and $\Delta\phi_\mathrm{BT}$ are the phase difference changes caused by environmental influences on the transmitter and receiver ends of AMZI, respectively. $\Delta\phi_\mathrm{ABT}$ is the phase difference between the two ends of AMZI itself.

For the LLO CV-QKD scheme, the main purpose of its proposal is to solve the security problem caused by the joint transmission of LO and signal light in the channel. The LO used in the LLO scheme is no longer generated by the laser at the transmitter side, but by an additional laser at the receiver side. In the original scheme, both LO and signal light are generated by the same laser with a uniform center frequency and phase, which can easily satisfy the interference condition, and even if a frequency drift occurs, it will affect both of them in the same way, so as not to cause a difference. However, in the LLO scheme, the LO and signal light from different lasers will have different frequencies and phases, affecting the effectiveness of interference-based coherent detection \cite{huang2015high}. During the operation of the system, the two lasers will each undergo a slow frequency drift, resulting in a change in the beat frequency, which increases the difficulty of signal demodulation. When the difference between the center frequencies of the two is too large and exceeds the bandwidth of the detector, it will result in the information loaded on the orthogonal components not being detected. Therefore, real-time feedback control of the center frequencies of the two lasers is also required to achieve long-term operation of the system \cite{huang2016continuous}. We take the LLO CV-QKD scheme with phase recovery by pilot as an example to analyze its phase error \cite{marie2017self}. The idea is to use the phase offset of the pilot to estimate the phase offset of the signal. However, an additional phase error is introduced in the estimation process
\begin{equation}
	\Delta\phi_\mathrm{LLO}=\Delta\phi_\mathrm{slow}+\Delta\phi_\mathrm{drift}+\Delta\phi_\mathrm{error},
\end{equation}
where $\Delta\phi_\mathrm{drift}$ is the phase error due to the independent phase drift of the two lasers in the LLO scheme in the middle of the time interval between the pilot and the signal light generation, and $\Delta\phi_\mathrm{error}$ is the phase measurement error due to measurements made on the pilot. Due to the time interval, $\Delta\phi_\mathrm{slow}$ is the error of the phase change between the pilot and neighboring signal light under the influence of environmental factors. Differences between lasers undoubtedly greatly reduce the stability of the system. 

In our DD CV-QKD scheme, the detection side uses direct detection without an interference structure. Firstly, we do not need to control the phase difference between the two optical fields of single photon interference, so the drift of temperature and other environmental factors will not affect the normal operation of the system. Secondly, we also no longer use coherent detection at the detection end, which eliminates the need for LO and avoids the problem of the difference of lasers in the LLO CV-QKD scheme. In addition, since the physical quantity measured by our scheme is the intensity of the optical field, the phase noise generated during channel transmission will be shielded in principle, eliminating the need for phase recovery using methods such as pilot \cite{marie2017self,qi2015generating,wang2018pilot}. Thus, our scheme does not introduce additional phase errors outside the quantum fluctuations of the signal light itself:
\begin{equation}
	\Delta\phi_\mathrm{DD}=0,
\end{equation}
By avoiding the use of interference structures, we have eliminated from the source the various problems mentioned earlier that could destabilize the system, which further improves the robustness of the system.

\begin{table*}[t]
	\renewcommand{\arraystretch}{1.5}
	\small % 将表格字体改为 small 或 footnotesize
	\setlength{\tabcolsep}{3pt} % 减小列与列之间的间距（默认通常是6pt）
	\begin{center}
		\caption{Phase error of QKD schemes.}
		\begin{tabular}{ccccc} 
			\hline
			\multirow{2}*{\textbf{Scheme}} & \multicolumn{3}{c}{\textbf{Sources of phase error}} & \multirow{2}*{\textbf{Total phase error}}\\
			& Environment & Device& Measurements &   \\ % 缩写表头文字
			\hline
			Phase-encoding DV-QKD & $\Delta\phi_\mathrm{A}+\Delta\phi_\mathrm{B}$ & $\Delta\phi_\mathrm{AB}$ & 0 & $\Delta\phi_\mathrm{A}+\Delta\phi_\mathrm{B}+\Delta\phi_\mathrm{AB}$ \\
			TLO CV-QKD & $\Delta\phi_\mathrm{AT}+\Delta\phi_\mathrm{BT}$ & $\Delta\phi_\mathrm{ABT}$ & 0 & $\Delta\phi_\mathrm{A}+\Delta\phi_\mathrm{B}+\Delta\phi_\mathrm{AB}$ \\
			LLO CV-QKD & $\Delta\phi_\mathrm{slow}$ & $\Delta\phi_\mathrm{drift}$ & $\Delta\phi_\mathrm{error}$ & $\Delta\phi_\mathrm{slow}+\Delta\phi_\mathrm{drift}+\Delta\phi_\mathrm{error}$ \\ % 如果公式太长，考虑简写
			DD CV-QKD & 0 & 0 & 0 & 0  \\
			\hline
		\end{tabular}
		\label{tab:robust}	
	\end{center}
\end{table*}

As shown in table \ref{tab:robust}, by avoiding the use of interference structures, we have eliminated from the source the various problems mentioned earlier that could destabilize the system, which further improves the robustness of the system. 

In the following, we analyse the robustness of the networks extended by these schemes. For the first three schemes, despite the existence of phase errors that can affect the robustness, the stable operation of the system can be guaranteed in a point-to-point system by adding a control system or using digital signal processing. However, when it is extended to a multi-user network, these robustness issues are further amplified.  

For phase-coded DV-QKD, the need for stable two pulses for single photon interference limits its network structure. Taking the simplest point-to-multipoint structure as an example, in the downstream structure, since the signals emitted by the QLT cannot be transmitted directionally to a particular user, all the QNUs have to ensure that their detectors are running all the time to receive randomly arriving quantum signals, which wastes a large portion of the detector bandwidth \cite{frohlich2013quantum}. In the upstream structure, each QNU transmits signals according to time-division multiplexing and shares the detectors at the QLT, which also reduces the communication efficiency of the whole network  \cite{hajomer2024continuous}.  Further, in the future quantum internet, it will contain important components such as routers, switches, repeaters, etc., but these nodes will inevitably suffer from signal loss during operation. As long as one pulse is lost, it will destroy the integrity of the two pulses in the phase-coded DV-QKD that perform single-photon interference, leading to interference failure and reducing the robustness of the network. All these problems become a major obstacle to its networking. 

For TLO CV-QKD, AMZI at the transmitter and receiver ends will cause random phase noise due to environmental factors. In addition, in a downlink structure, when the number of users increases, in order to ensure that coherent detection at the user end can reach the shot noise limit, the transmitter end needs to prepare a higher-power LO, which will increase the leakage noise of the LO. 

For LLO CV-QKD, when forming a multi-user network, each node needs to be equipped with a tunable laser. Any two parties that want to share the key have to ensure that their lasers have the same centre frequency. This results in the need for an overall control system for the entire network to regulate all the lasers in the network in real time, which is very difficult and limits the ability of the network to operate stably over a long period of time. 

In addition, after these schemes form a network, the different environments where each user is located can affect the interference structure, which in turn leads to exponentially increasing and uncontrollable errors caused by environmental factors, threatening the robustness of the network. These robustness problems are caused by the harsh interference conditions in the interference structure. 

When our DD CV-QKD scheme is extended to a network, the avoidance of the interference structure can solve these problems in principle and thus achieve higher robustness. Firstly, benefiting from the advantages of coherent state signals, we have the flexibility to choose either upstream or downstream structure and generate keys with all users simultaneously in the downstream structure. Subsequently, the interference-free structure allows us not to control harsh interference conditions, and fluctuations in environmental factors do not have an impact on the robustness of the system. These advantages are even more evident in networks, as there is no need for any real-time control of each node. In addition, the direct detection method is very similar to that of current classical optical communication networks, where this method is well compatible with structures such as routers, switches, repeaters, and so on. It further proves the great potential of our scheme for the future quantum internet.

\begin{figure}[!t]
	\centering
	\subfigure[]{\includegraphics[width=8cm]{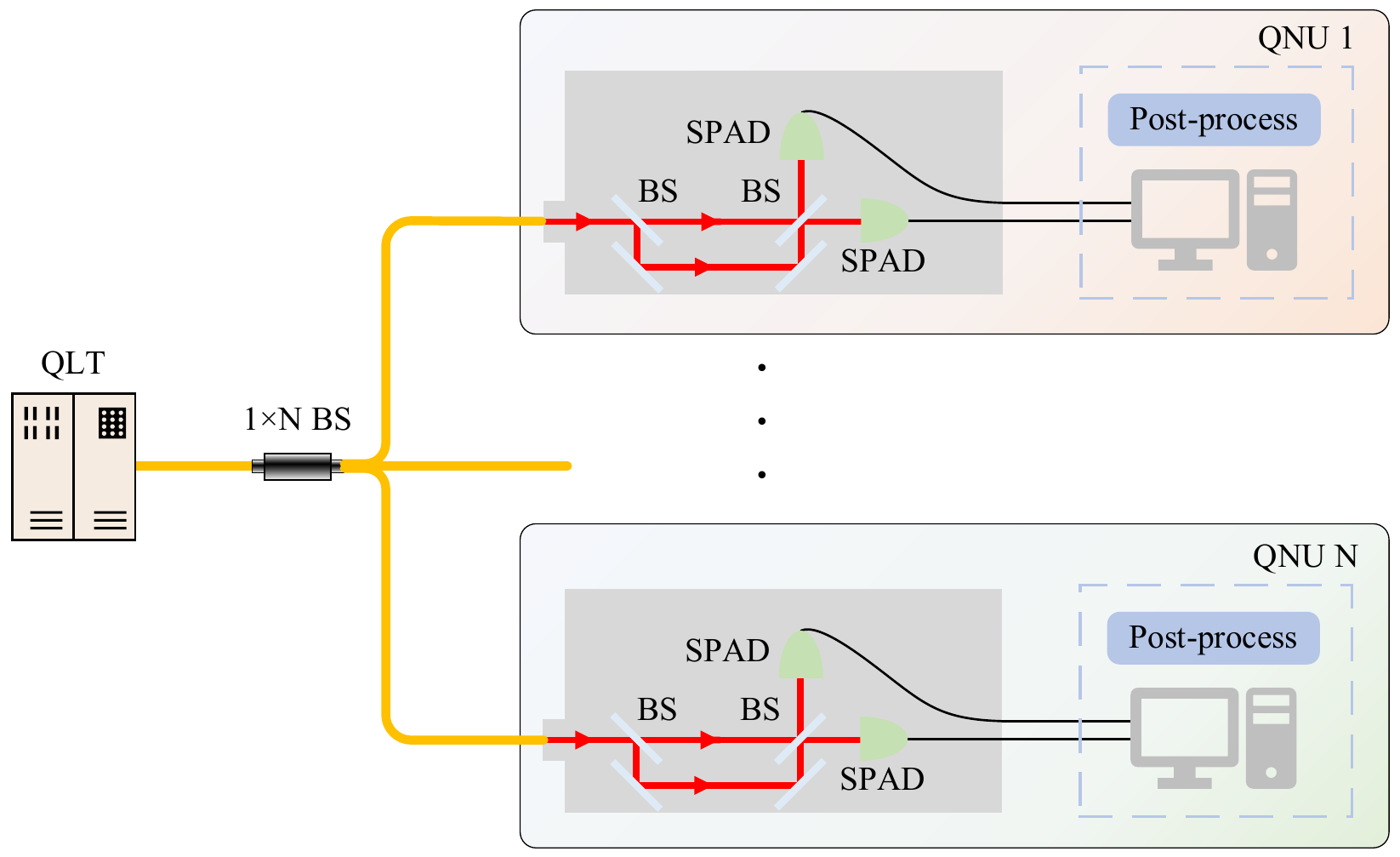} \label{SPAD}}
	
	\centering
	\subfigure[]{\includegraphics[width=8cm]{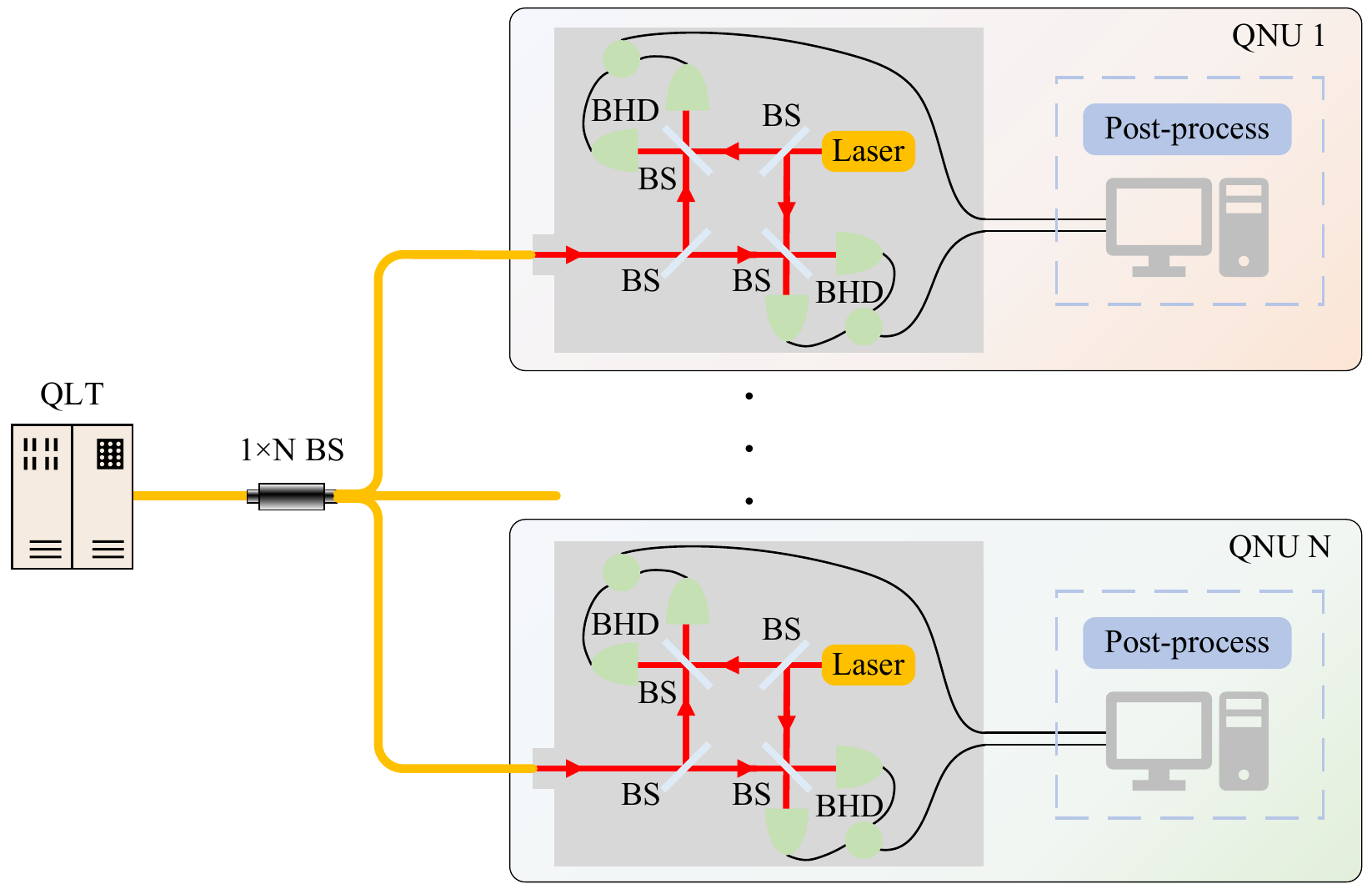} \label{BHD}}
	
	\centering
	\subfigure[]{\includegraphics[width=8cm]{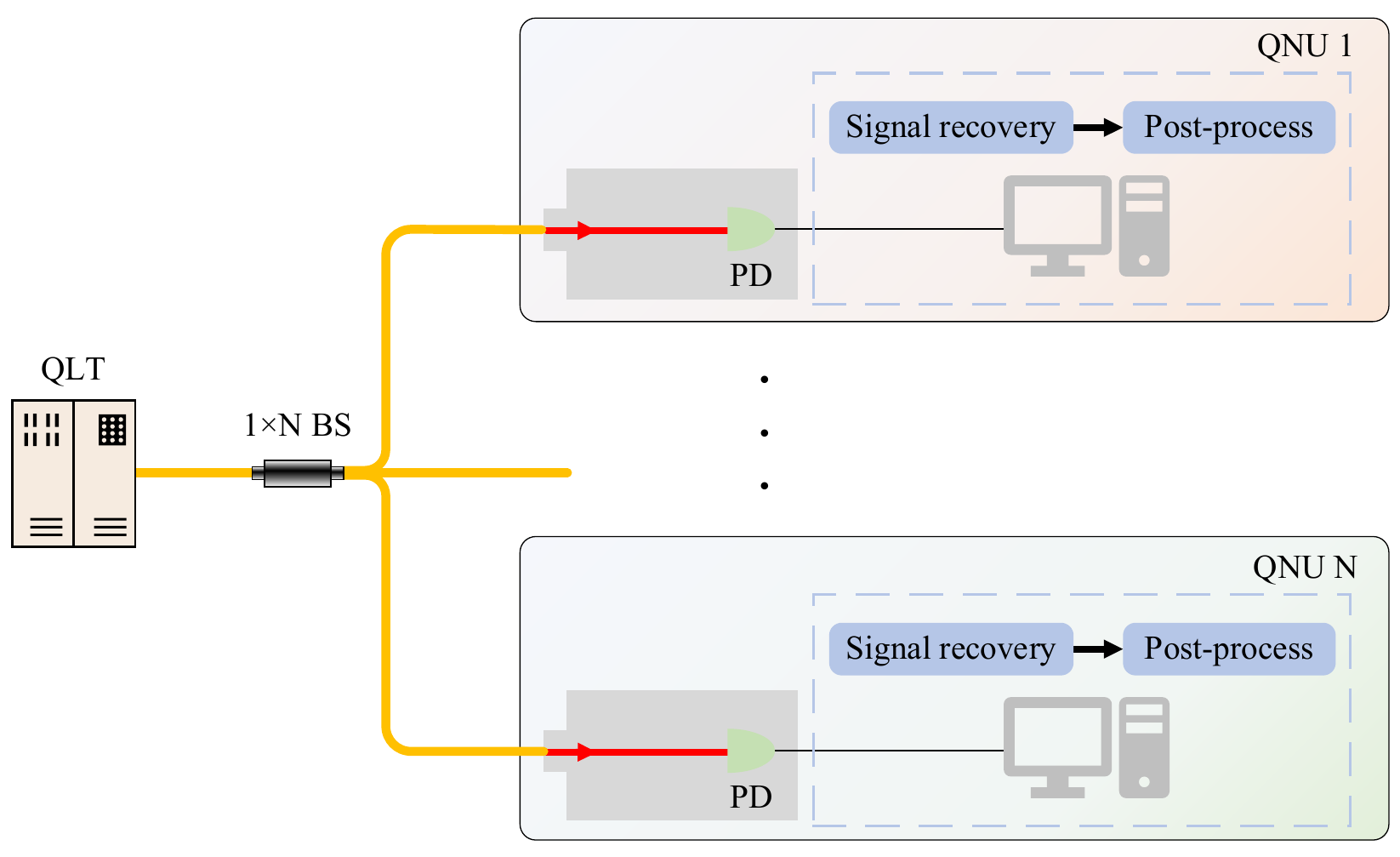} \label{PD}}
	\caption{QNU structure in downstream QAN for three detection methods (QNU, quantum network unit; QLT, quantum line terminal; BS, beam splitter; HWP, half-wave plate; SPAD, single photon avalanche diode;PBS, polarization beam splitter; BHD, balanced homodyne detector; LO, local oscillator; PD, photodetector). (a) The single-photon detection QNU structure in downstream QAN using SPAD. (b) The coherent detection QNU structure in downstream QAN using BHD. (c) The direct detection QNU structure in downstream QAN using PD.}
	\label{fig:detection}
\end{figure}

\section{Supplementary Note 7: QAN cost-effectiveness analysis}

We further demonstrate the cost-effectiveness of the DD CV-QAN scheme. We compare the cost of the existing downstream DV-QAN and CV-QAN with our DD CV-QAN. For QLT, the tunable laser is the main overhead as a coherent state source for CV-QAN and a weak coherent state source for the equivalent single photon source in DV-QAN. In the following analysis for QLT, we consider only the cost of the tunable laser. For QNU, the passive optics and the data processor used in each scheme are similar, so we focus on the detector cost of each scheme and the cost of the tunable laser as the LO in the CV-QAN scheme.

For the optics that may be involved in the costing, we set up a cost relationship with reference to the prices of Thorlabs \cite{thorlabs}, a widely used optics manufacturer. We use the cost of PD as the unit cost and denote the cost of PD as $C_{\mathrm{PD}}$, the cost of BHD as $4\ C_{\mathrm{PD}}$, the cost of SPAD as $10\ C_{\mathrm{PD}}$, and the cost of tunable laser as $20\ C_{\mathrm{PD}}$.  Based on the types and numbers of devices included in QLT and QNU in each scheme, we will calculate the total cost $P$ required for a downstream QAN containing $N$ users.

For a downstream DV-QAN using single-photon detection, we take a phase-encoding system as an example, as shown in figure \ref{fig:detection} \subref{SPAD}. The QLT side requires a tunable laser as a weakly coherent state source equivalent to a single-photon source. Two SPADs are required on the QNU side to measure the two outputs of single photon interference. Then the total cost of the downstream QAN for its N-users is
\begin{equation}
	P_{\mathrm{DV}}=20\ C_{\mathrm{PD}}+2N\cdot10\ C_{\mathrm{PD}}=(20+20N)\ C_{\mathrm{PD}}.
\end{equation}

For the downstream CV-QAN using coherent detection, we take the LLO heterodyne detection system as an example, as shown in figure \ref{fig:detection} \subref{BHD}. A tunable laser is needed on the QLT side as the coherent-state source. Two BHDs are needed on the QNU side to measure the two quadrature components of the light field respectively, in addition to a tunable laser as the LO for the LLO scheme. Then the total cost of its N-user downstream QAN is
\begin{equation}
	P_{\mathrm{LLO}}=(N+1)\cdot 20\ C_{\mathrm{PD}}+2N\cdot4\ C_{\mathrm{PD}}=(20+28N)\ C_{\mathrm{PD}}.
\end{equation}
The TLO scheme is similar to the LLO scheme, but there is no additional laser at the receiving end.
\begin{equation}
	P_{\mathrm{TLO}}=20\ C_{\mathrm{PD}}+2N\cdot4\ C_{\mathrm{PD}}=(20+8N)\ C_{\mathrm{PD}}.
\end{equation}

For our proposed downstream DD CV-QAN, a tunable laser is also required on the QLT side as the coherent state source. And only one PD is needed on the QNU side to complete the detection of coherent state and no additional tunable laser is needed as the source of the LO, as shown in figure \ref{fig:detection} \subref{PD}. Then the total cost of its N-user downstream QAN is
\begin{equation}
	P_{\mathrm{DD}}=20\ C_{\mathrm{PD}}+N\cdot C_{\mathrm{PD}}=(20+N)\ C_{\mathrm{PD}}.
\end{equation}

\begin{table*}[t]
	\renewcommand{\arraystretch}{1.5}
	\begin{center}
		
		\caption{Total cost of N-user downstream QANs.}
		\resizebox{\linewidth}{!}{
			\begin{tabular}{cccccccc} % <-- Alignments: 1st column left, 2nd middle and 3rd right, with vertical lines in between
				\hline
				\multirow{2}*{\textbf{Scheme}} & \textbf{QLT cost} & \multicolumn{4}{c}{\textbf{QNU cost}} & \multirow{2}*{\textbf{Total cost}}& \multirow{2}*{\textbf{Security}}\\
				
				&Tunable laser & SPAD & BHD & PD & Tunable laser  \\
				\hline
				Phase-encoding DV-QAN & $20\ C_{\mathrm{PD}}$ & $2N\cdot 10\ C_{\mathrm{PD}}$ & 0 & 0 & 0 & $(20+20N)\ C_{\mathrm{PD}}$ &/  \\
				TLO CV-QAN & $20\ C_{\mathrm{PD}}$ & 0 & $2N\cdot 4\ C_{\mathrm{PD}}$ & 0 & $NC_{\mathrm{PD}}$ & $(20+8N)\ C_{\mathrm{PD}}$& with LO vulnerability  \\
				LLO CV-QAN & $20\ C_{\mathrm{PD}}$ & 0 & $2N\cdot 4\ C_{\mathrm{PD}}$ & 0 & $N\cdot 20\ C_{\mathrm{PD}}$ & $(20+28N)\ C_{\mathrm{PD}}$&without LO vulnerability \\
				DD CV-QAN & $20\ C_{\mathrm{PD}}$ & 0 & 0 & $N\cdot C_{\mathrm{PD}}$ & 0 & $(20+N)\ C_{\mathrm{PD}} $&with $A$ vulnerability  \\
				\hline
				
			\end{tabular}
		}
		\label{tab:cost}	
	\end{center}
\end{table*}

\begin{figure}[!h]
	\centering
	\includegraphics[width=8.5cm]{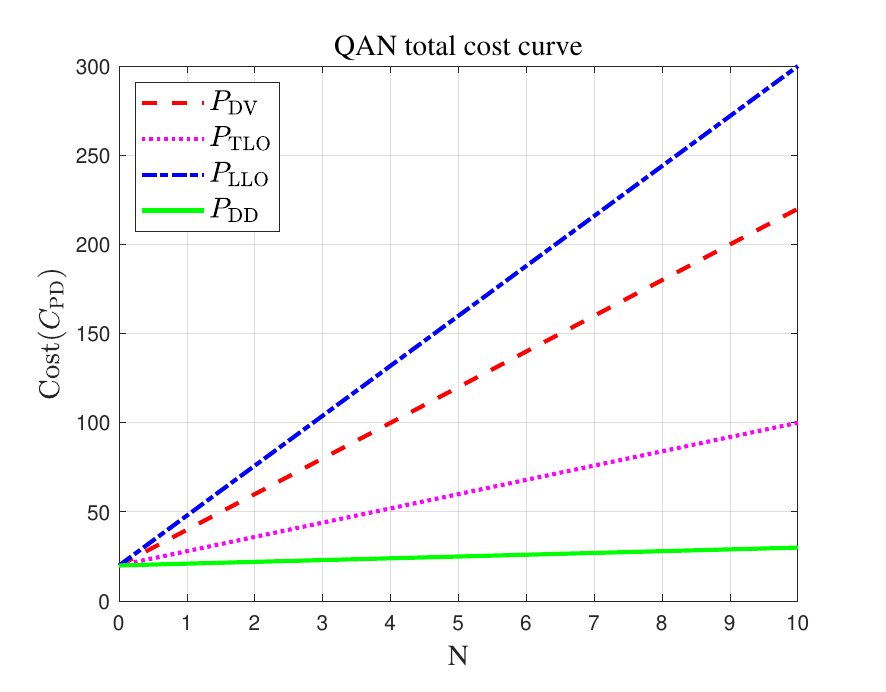}
	\caption{Total cost curve of QANs for three detection methods.The red dashed line indicates single-photon detection DV-QAN, the blue dash-dot line indicates coherent detection CV-QAN, and the green solid line indicates DD CV-QAN. }
	\label{fig:cost}
\end{figure}

Table \ref{tab:cost} demonstrates the components of four QAN costs. We further plot the total cost curves of the four QANs versus the number of users $N$. From figure \ref{fig:cost}, we can see that the total cost of other schemes increases rapidly as the number of users $N$ increases. In our scheme, on the other hand, since the cost of the detector PD used is significantly lower than that of SPAD and BHD, the cost of QNU is very low, and the total cost of QAN rises extremely slowly as the number of users increases. In addition, although we compare here only with downstream structural QANs, the cost of QNUs for upstream structural QANs is still higher than that in DD CV-QAN due to the presence of lasers. The cost-effectiveness of our scheme is demonstrated by its significant cost advantage over other schemes, especially in the case of a large number of users, making this scheme more suitable for large-scale deployment.

It should be noted that, as analyzed in the section on practical security above, our scheme is somewhat similar to the TLO scheme and also carries the DC component $A$ vulnerability similar to the LO vulnerability. Therefore, the practical security is indeed weaker compared to the LLO scheme. However, even without considering lasers, our detector is still cost-effective compared to the TLO scheme.

Furthermore, based on our analysis in \textbf{supplementary note 5}, compared with coherent detection CV-QKD, our scheme cannot achieve the GHz level of electronic bandwidth of AC amplifiers used in coherent detection because DC-coupled amplifiers must be used at the PD backend. This limits our signal repetition frequency, resulting in a lower SKR.

\section{Supplementary Note 8: The method for calibrating the DC component at the receiver}

In the process of transforming the recovered minimum-phase signal back into the original Gaussian modulated signal, we estimate the magnitude of the DC component arriving at the receiver by taking the average of the real part of the minimum-phase signal. We can divide the signal into two parts: the DC component part and the Gaussian modulated signal part after spectral shifting. Within a single data frame, the DC component is identical and can be calculated statistically. The Gaussian-modulated signal, however, is unique for each frame and does not provide eavesdroppers with an opportunity to access a copy. Additionally, the statistical mean of the Gaussian-modulated signal is zero, which does not affect the calculation of the DC component. We will further introduce the method for calibrating the DC component at the receiver.

The minimum-phase signal recovered at the receiver is
\begin{equation}
	\hat E_{MPr}(t)=A_r+(\hat a_{sr}(t)+\hat a_{i}(t))\exp(i\omega_{IF}t).
\end{equation}
Under ideal conditions, $A_r$ remains constant within a data frame, and the expectation of $\hat a_{sr}(t)+\hat a_{i}(t)$ is 0. Therefore, we can calibrate $A_r$ as follows:
\begin{equation}
	A_r=E[\Re(\hat E_{MPr}(t))],
\end{equation}
where $\Re(x)$ denotes the real part of the $x$. During the experiment, the expected value will be calculated by taking the average. Here, fluctuations in the DC component caused by system factors such as laser jitter have been attributed to shot noise. Statistical errors caused by small data volumes will result in excess noise.

\section{Supplementary Note 9: Shot noise of DD CV-QKD}
In the theoretical security analysis, we derived the detection operators for DD CV-QKD and heterodyne detection CV-QKD after shot noise normalization. Here, we further analyze the shot noise calibration and normalization process in actual systems.

We first analyze heterodyne detection CV-QKD. In actual systems, the measurement results of coherent detection contain a constant coefficient $R_c$ \cite{kikuchi2015fundamentals}:
\begin{equation}
	\begin{aligned}
		&\hat x_{het-act}=\frac{R_c}{\sqrt2}(\hat x_{a_{sr}(t)}+\hat x_{a_i(t)}),
		\\&\hat p_{het-act}=\frac{R_c}{\sqrt2}(\hat p_{a_{sr}(t)}+\hat p_{a_i(t)}),
	\end{aligned}
\end{equation}
where $R_c \propto \sqrt{P_l}$, $P_l$ is the power of LO. The shot noise unit when there is no signal input is
\begin{equation}
	\begin{aligned}
		&\langle\Delta\hat x^2_{het-SNU-act}\rangle=R_c^2\langle\Delta\hat  x^2_{a_i(t)}\rangle=R_c^2V_{vac},
		\\&\langle\Delta\hat p^2_{het-SNU-act}\rangle=R_c^2\langle\Delta\hat  p^2_{a_i(t)}\rangle=R_c^2V_{vac}.
	\end{aligned}
\end{equation}
It can be seen that the calibrated value of shot noise in the actual system is proportional to $R_c$, that is, proportional to $P_l$. The greater the LO power, the greater the shot noise. Perform shot noise normalization to obtain
\begin{equation}
	\begin{aligned}
		&\hat x_{het-act}\leftarrow\frac{\hat x_{het-act}}{R_c\sqrt{V_{vac}}}
		=\frac{\hat x_{a_{sr}(t)}+\hat x_{a_i(t)}}{\sqrt{2V_{vac}}},
		\\&\hat p_{het-act}\leftarrow\frac{\hat p_{het-act}}{R_c\sqrt{V_{vac}}}
		=\frac{\hat p_{a_{sr}(t)}+\hat p_{a_i(t)}}{\sqrt{2V_{vac}}}.
	\end{aligned}
\end{equation}
Since $R_c$ is canceled out, the normalized operator is consistent with the form in the theoretical analysis and is independent of LO power.

In the DD CV-QKD scheme, the DC component is similar to LO. The effect of the DC component is reflected in the minimum phase signal recovered at the receiver:
\begin{equation}
	\begin{aligned}
		\hat {E’}_{MPr}(t)&=\sqrt{\frac{\hat I(t)}{\mu}}\exp(i\hat\phi_{E’}(t))
		\\&=A_r+(\hat a_{sr}(t)+\hat a_i(t))\exp(i\omega_{IF}t).
	\end{aligned}
\end{equation}
Further restoration of the signals light field gets
\begin{equation}
	\hat a_{sr}(t)+\hat a_i(t)=(\hat {E’}_{MPr}(t)-A_r)\exp(-i\omega_{IF}t).
\end{equation}
It can be seen that at this point, we have removed the influence of the DC component $A_r$, and the measurement results are independent of $A_r$. Therefore, in the actual calibration of shot noise, the size of the calibration value is unrelated to $A_r$, which is different from heterodyne detection CV-QKD.

Although we still ensure that the shot noise calibration is the same as $A_r$ during system operation in our experiment, this is mainly to ensure the accuracy of the electronic noise calibration, rather than to ensure the accuracy of the shot noise calibration as in heterodyne detection CV-QKD. According to the analysis in \textbf{supplementary note 10}, the electronic noise calibration value changes with $A_r$.

\begin{figure}[!t]
	\centering
	\includegraphics[width=9cm]{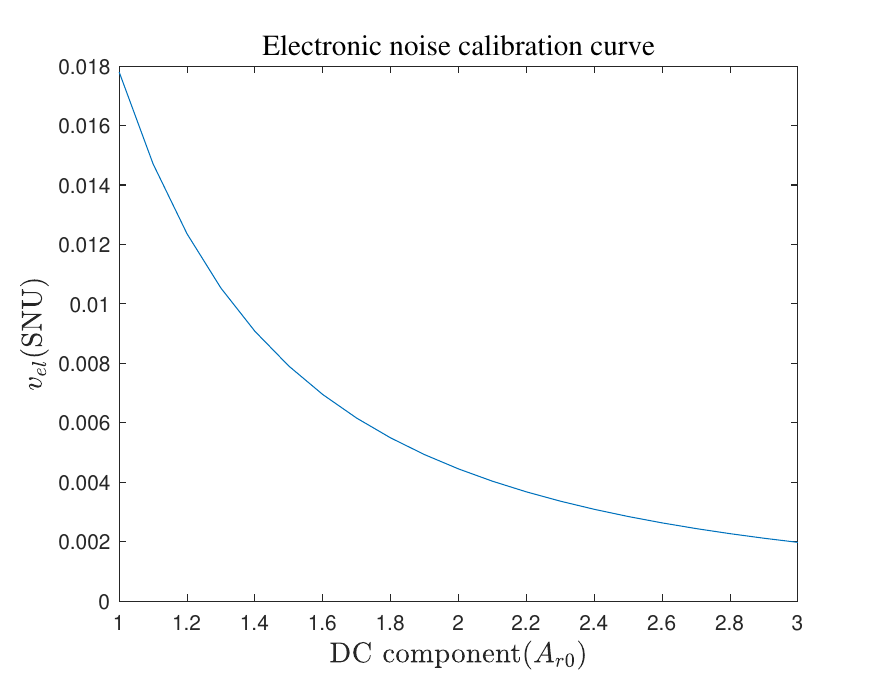}
	\caption{Relationship between electronic noise calibration and DC component.}
	\label{fig:velcc}
\end{figure}

\section{Supplementary Note 10: Detection noise of DD CV-QKD}

According to theoretical security derivations, the detection noise of our DD CV-QKD scheme has the same form as that of heterodyne detection:
\begin{equation}
	\chi_{DD}=\chi_{het}=[1+(1-\eta)+2v_{el}]/\eta.
\end{equation}
As we analyze in supplementary note 5, our scheme differs from the heterodyne detection scheme in terms of electronic noise calibration. In the digital signal processing process that is the same as the signal, the noise added to the photocurrent will be amplified and superimposed on the two quadrature components, becoming the electronic noise we calibrated. It should be noted that the electronic noise calibrated in our scheme decreases as the DC component $A_r$ at the receiver increases, rather than remaining constant as in heterodyne detection. The main reason is that the Kramers-Kronig relation includes nonlinear operations such as square root and logarithmic operations, and the derivatives of these functions decrease as the independent variable (i.e., photocurrent) increases. Therefore, when fluctuations with the same variance are superimposed on different photocurrent values, the variance of the calculation results will also be different. Therefore, when $A_r$ increases, the photocurrent increases, and the variance of the same noise superimposed on it will decrease after calculation, resulting in a change in the electronic noise calibration value.

We used the electronic noise data calibrated in the experiment and changed the size of $A_r$. Through Monte Carlo simulation, we plotted a curve showing how the electronic noise calibration value $v_{el}$ changes with $A_r$, as shown in figure \ref{fig:velcc}. In the simulation process, we used the DC component value $A_{r0}$ measured in the experiment as the unit. From $A_{r0}$ to $3A_{r0}$, $v_{el}$ continued to decrease.

During the experiment, we calibrated the electronic noise to be two orders of magnitude smaller than the detection noise. For the detection noise as a whole, the tiny changes in electronic noise have a very limited impact on the overall result, so we can also assume that the detection noise is constant.